\documentclass[reqno]{article}
\usepackage{authblk}
\usepackage{amssymb,latexsym,amsmath,amsfonts}
\usepackage{graphicx}
\usepackage{color,enumerate}
\usepackage{amsmath,amsfonts,amssymb}
\usepackage[mathscr]{eucal}
\usepackage{bbm}
\usepackage{booktabs}
\usepackage{float}
\usepackage{hyperref}
\usepackage{amsthm}
\usepackage[mathscr]{eucal}
\usepackage{paralist}
\usepackage{rotating}
\usepackage{multirow}
\usepackage[left=1.25in, right=1.25in,bottom=1.25in]{geometry}
\newtheorem{theorem}{Theorem}

\newtheorem{proposition}{Proposition}

\theoremstyle{definition}
\newtheorem{definition}[theorem]{Definition}

\theoremstyle{remark}
\newtheorem{remark}{Remark}

\newcommand{\RN}[1]{\uppercase\expandafter{\romannumeral#1}}

\newcommand{\edit}[1]{\textcolor{black}{#1}}

\providecommand{\keywords}[1]{\textbf{Keywords: } #1}

\providecommand{\ams}[1]{\textbf{Mathematics Subject Classification(s): } #1}

\newcommand\extrafootertext[1]{%
	\bgroup
	\renewcommand\thefootnote{\fnsymbol{footnote}}%
	\renewcommand\thempfootnote{\fnsymbol{mpfootnote}}%
	\footnotetext[0]{#1}%
	\egroup
}

\newcommand\correspondingauthor{\thanks{Corresponding author.}}
\title{Pattern Formation in Mesic Savannas}
\author[1,2]{Denis D. Patterson\correspondingauthor}
\affil[1]{High Meadows Environmental Institute, Princeton University, Princeton, NJ (denispatterson@princeton.edu)}
\author[2]{Simon A. Levin}
\affil[2]{Department of Ecology and Evolutionary Biology, Princeton University, Princeton, NJ (slevin@princeton.edu)}
\author[3]{ A. Carla Staver}
\affil[3]{Department of Ecology and Evolutionary Biology, Yale University, New Haven, CT (carla.staver@yale.edu)}
\author[4,5]{Jonathan D. Touboul}
\affil[4]{ Department of Mathematics, Brandeis University, Waltham MA  (jtouboul@brandeis.edu)}
\affil[5]{Volen Centre for Complex Systems, Brandeis University, Waltham MA}

\date{\today}
\begin{document}

\maketitle

\begin{abstract}
We analyze a spatially extended version of a well-known model of forest-savanna dynamics, which presents as a system of nonlinear partial integro-differential equations, and study necessary conditions for pattern-forming bifurcations. Analytically, we show that homogeneous solutions dominate the dynamics of the standard forest-savanna model, regardless of the length scales of the various spatial processes considered. However, several different pattern-forming scenarios are possible upon including spatial resource limitation, such as competition for water, soil nutrients, or herbivory effects. Using numerical simulations and continuation, we study the nature of the resulting patterns as a function of system parameters and length scales, uncovering subcritical pattern-forming bifurcations and observing significant regions of multistability for realistic parameter regimes. Finally, we discuss our results in the context of extant savanna-forest modeling efforts and highlight ongoing challenges in building a unifying mathematical model for savannas across different rainfall levels.
\end{abstract} 
\vspace*{10pt}

\vspace*{10pt}

\ams{45K05, 35B36, 92F05}

\extrafootertext{\noindent This paper was inspired to a large extent by Mayan Mimura’s foundational mathematical work on pattern formation, which had such influence on the understanding of the dynamics of ecological systems. It is our pleasure to dedicate this paper to Prof. Mimura's memory. \\[3pt] Simon Levin and Denis Patterson thank the NSF for support via the grant DMS-1951358, and Simon Levin also appreciates support from the Army Research Office Grant W911NF-18-1-0325. Carla Staver appreciates support from NSF Grant DMS-1951394 and Jonathan Touboul appreciates support from NSF Grant DMS-1951369.}

\newpage

\section{Introduction}
Savanna and tropical forest are two of the most crucial biomes with regard to current conservation efforts, with savannas covering approximately one eight of the Earths total landmass and tropical forests a key carbon sink. Savannas are typically defined by coexistence of grass and trees with an open canopy. Their emergence has proven notoriously difficult to model and predict and relies on multiple complex processes, including fire disturbance and herbivory~\cite{van2003effects}.  Savannas are also diverse in nature; they exist over a wide range of mean annual rainfall (MAR), considered to be a strong determinant of vegetative cover, with the driest savanna found below 1000mm MAR and the wettest occurring above 2000mm MAR. Recent empirical findings suggest that savanna and tropical forest constitute stable alternatives at intermediate rainfall levels~\cite{staal2020hysteresis,staver2011tree,staver2011global}. This bistability suggests the possibility of switching or even irreversibly tipping between these states. Hence there is intense interest in developing appropriate models to capture this bistability, and thereby properly evaluate the stability and resilience to perturbations of tropical forest-savanna ecosystems.

There a burgeoning literature regarding the mathematical modeling of savannas and adjacent biomes such as woodland and tropical forest. Typically, these models involve interactions between one or more species of trees, shrubs and grasses, with each functional type of vegetation having different tolerances or reactions to fire and herbivory~\cite{staver2020seasonal}. Some of these models of savannas are spatially implicit and describe only proportions of a landscape occupied by each cover type~\cite{hoyer2021impulsive,staver_levin_2012}. These models are accurate for describing well-mixed regions of vegetation, which is typically true at small length scales. However, when considering larger-scale systems that may include multiple types of vegetation, frameworks with explicit spatial dimensions are required. In recent years, such models were proposed either relying on partial differential equations (PDEs)~\cite{champneys2021bistability,wuyts2019tropical}, cellular automata~\cite{hebert2018edge,wuyts2022emergence} or spatial stochastic models~\cite{durrett2018heterogeneous,durrett2015coexistence}. Much of this work has focused on the ranges and mechanisms for coexistence or multistability between biomes such as grasslands, savanna, woodland or forest in the tropics.

In spite of the significant modeling effort outlined above, there has been somewhat less attention paid to the spatial structure within the ecosystems predicted by these mathematical models, at least for mesic savannas, i.e. those at intermediate to high rainfall. In contrast, there is an extensive literature regarding pattern formation in vegetation models of dryland and semi-arid ecosystems, with the tiger bush model being among the best known examples~\cite{lefever1997origin}. This work principally considers low rainfall ecosystems (less than 500mm MAR) in which water limitations highly constrain vegetative growth, and hence the models necessarily involve explicitly hydrological modeling of ground and soil water content, in addition to the vegetation state variables~\cite{kefi2007spatial,meron2012pattern2,von2001diversity,siero2015striped}. Despite their more arid setting, the majority of these models share the qualitative feature of bistability with forest-savanna models, although in this case the bistability is between a state with non-zero vegetative biomass and a bare ground state. One of the key lessons from this body of work is that spatial patterning can have significant implications for the stability of ecosystems and may even qualitatively change the types of transitions that occur between alternative stable states. In particular, pattern-forming instabilities in a homogeneous stable state may not be a precursor to ecosystem collapse, but may instead allow the system to maintain a state of higher vegetative biomass for a much wider region of parameter space~\cite{rietkerk2021evasion}. It is evidently important to understand the extent to which these conclusions from pattern formation in arid ecosystems carry over to forest-savanna systems at intermediate rainfall, which have very different dominant interactions with different (relative) spatial scales.

Various empirical studies have reported evidence for spatial patterning in savanna ecosystems, particularly in nutrient deficient savannas~\cite{lejeune2002localized} and at the boundaries between savanna and forest systems~\cite{rietkerk2008regular}. This work inspired several branches of mathematical research seeking to explain these observations with different patterning mechanisms proposed for dry versus wetter savannas. Many models of spatial patterning at lower MAR are adapted from the arid ecosystem modeling paradigms and focus on patterns emerging from water resource competition and scarcity~\cite{baudena2013complexity,groen2017spatially,tzuk2020role}. These models typically involve systems of PDEs (sometimes with nonlocal operators for seed dispersal) with state variables for woody and herbaceous cover, in addition to soil and groundwater quantities. They produce a wide array of patterned steady states via Turing-type bifurcations. In addition, numerical continuation has proven an invaluable tool in this work, allowing detailed descriptions of the dynamics away from local bifurcations and thus highlighting possible paths for regime shifts. However, these works typically do not emphasize fire or herbivory effects, which are likely to be more significant at intermediate to higher MAR, and appear more suited to study ecosystem transitions and stability at the boundary between savanna and semi arid states with low vegetative cover. Other researchers have proposed spatially explicit mathematical models of mesic savannas focused on disturbance and facilitative interactions appropriate for higher MAR settings that can similarly produce patterned steady states~\cite{lejeune2002localized,martinez2013spatial,tega2022spatio}. These works serve as an important proof of principle in terms of the mechanisms which can lead to stable heterogeneous savanna ecosystems, but have certain drawbacks that limit their applicability as general models for studying transitions between savanna and forest states in the bistable MAR range. For instance, some of these models are posed with a single state variable for biomass~\cite{lejeune2002localized,martinez2013spatial}, affording mathematical tractability, but limiting their ability to reflect the functional properties of ecosystems by discriminating between different cover types. Other models require quite specific spatial kernels to establish stable patterns, raising questions about generality and robustness~\cite{tega2022spatio}.

In this work, we aim at an intermediate complexity description of mesic savannas using a spatially explicit nonlocal mathematical model based on a well-studied forest-savanna framework often referred to as the Staver-Levin model~\cite{durrett2015coexistence,staver_levin_2012}. We demonstrate that stable spatially patterned ecosystems are possible in this framework under realistic assumptions on the lengthscales of the dominant spatial interactions at play. In order to retain a relatively tractable mathematical model, we don't explicit add hydrological constraints to our model, but we do include a nonlocal resource limitation on forest tree growth, in addition to the spatial fire spread and seed dispersal processes, since this turns out to be crucial for the formation of stable heterogeneous solutions. A particular advantage of our framework is that we can capture transitions between different biomes in much greater detail than previous modeling efforts in this domain. For example, our model draws clear distinctions between grassland, woodland, savanna and closed canopy tropical forest ecosystems, making it appropriate for more detailed future studies of ecosystem resilience and transitions between these, potentially multistable, alternative states. Moreover, we model spatial interactions via nonlocal operators, which is considered particularly appropriate for potentially long-range processes such as seed dispersal~\cite{nathan2012dispersal,thompson2008plant}, affording us considerable flexibility to adapt our model to different scenarios and empirical data, as well as giving us robust and general qualitative conclusions.

The paper is organized as follows: In Section \ref{sec_model_overview}, we introduce a spatially extended version of the Staver-Levin model with \emph{nonlocal} spatial interactions between the various vegetation types. We then perform a linear stability analysis to show that the steady-states of the original nonspatial model remain stable in the spatially extended model under very mild assumptions. In Section \ref{sec_resource_limitation}, we introduce a new version of the Staver-Levin model with resource limitation on the growth of forest trees and show that in this more realistic model homogeneous steady-state solutions may lose stability via Turing bifurcations, leading to the emergence of heterogeneous steady states. We investigate the structure of the bifurcations and the emerging heterogeneous solutions in detail for the grass-forest model as a function of system parameters and the different length scales in the model. We further show that in the full resource limited spatial Staver-Levin model, linear stability analysis once more confirms the presence of Turing bifurcations and the emergence of heterogeneous steady states is confirmed by numerical simulations of the system. Section \ref{sec_conclusions} provides a discussion of the main results and their implications for tropical forest-savanna ecosystems and their management.

\section{A spatially extended forest-savanna model}\label{sec_model_overview}
\subsection{Model  overview}
The Staver-Levin model (SL model) describes the interactions between savanna trees \edit{(T), savanna tree saplings (S),} forest trees \edit{(F)}, and grass \edit{(G)}, and was originally motivated by empirical evidence supporting forest-savanna bistability at intermediate rainfall in the tropics~\cite{staver2011global}. \edit{In particular, there is evidence that frequent fires can prevent the formation of closed canopy forests and help to maintain savannas in regions that would otherwise be suitable for forest establishment based on their climate~\cite{bond2008limits}.}

\edit{The original SL model~\cite{staver2011tree,staver_levin_2012} incorporates spatial extent implicitly by assuming that all vegetation types are spatially well-mixed and hence it tracks the proportion of space occupied by each vegetation type. In this framework, grass represents an ``open'' patch on which new trees can grow, but grass patches also carry fires that limit the expansion of both savanna and forest trees, albeit via different mechanisms. Forest trees can grow on grass, sapling or savanna occupied patches, at a rate associated with the amount of seeds available on that patch coming from adjacent forest trees. Similarly, savanna tree saplings grow on grass patches at a rate depending on the prevalence of adult savanna trees in their vicinity. Grass represents the primary flammable cover that carries fires which kill forest trees. Forest tree mortality to fire thus depends on the level of grassy cover and this mortality (burning) rate is denoted by $\phi(G)$. Empirically, fire is very frequently observed in savanna systems below approximately $40\%$ tree cover, but is very rarely observed in savanna systems with more than $40\%$ tree cover~\cite{staver2011tree}. Moreover, percolation models of fire spread dynamics support a sharp threshold for fire activation as a function of flammable cover~\cite{schertzer2015implications}. Hence the forest tree burning rate, $\phi(G)$, depends on grass levels and is typically chosen to be an increasing sigmoidal function with a sharp transition from low-fire to fire-prone regimes.}

\edit{Fire carried by the grassy layer also impacts the maturation of savanna saplings to adult savanna trees. However, savanna saplings are ``top killed'', rather than totally destroyed, by fire and their recruitment to adult savanna trees is merely delayed as they are able to resprout later. Thus the sapling-to-savanna recruitment rate, $\omega(G)$, depends on grass levels, and is chosen to be a decreasing sigmoidal function with an abrupt transition from high recruitment in the absence of fire to low recruitment in the fire-prone regime. Adult savanna trees are adapted to fire and do not suffer significant excess mortality in a fire-prone environment. Finally, since grass grows much more quickly than the other vegetation types, we assume that whenever savanna trees, savanna saplings or forest trees die, either via natural mortality or fire, they immediately revert back to the grass state.} 

\edit{The interaction rules  described above generate the following system of ODEs for the \emph{proportions of space} covered by each of the four functional types of vegetation:}
\begin{subequations}\label{SL_ode}
	\begin{alignat}{2}
		\dot{G} &= \mu S + \nu T + \phi(G)F - \alpha G F - \beta G T, \\
		\dot{S} &= -\mu S - \omega(G) S - \alpha S F + \beta G T,\\
		\dot{T} &= -\nu T + \omega(G) S - \alpha T F, \\
		\dot{F} &= \alpha (G + S + T)F - \phi(G)F, \\
		1 &= G + S + T + F,
	\end{alignat}
\end{subequations}
where $G$ denotes the grass cover proportion, $S$ is the savanna saplings proportion, $T$ is the savanna tree cover proportion and $F$ is the forest tree cover proportion. \edit{The positive constants $\mu$ and $\nu$ are the mortality rates of saplings and savanna trees respectively, while $\alpha$ is the forest tree birth rate and $\beta$ the savanna tree birth rate. The algebraic constraint given by the final equation in \eqref{SL_ode} ensures that all space is filled by one of the four types of vegetation. Table \ref{table.SL_parameters} summarizes the parameters of the SL model, along with their ecological interpretations and default numerical values.  In addition to the four-functional-type model described above, numerous other related models with the similar underlying interaction rules, some including explicit spatial extent, have been studied in the  literature~\cite{durrett2015coexistence,durrett2018heterogeneous,hoyer2021impulsive,li2019spatial,schertzer2015implications,touboul2018complex,wuyts2019tropical}.} 

\edit{A key feature of the SL model is the threshold response to fire as a function of flammable cover. The \edit{nonlinear} functions $\phi$ and $\omega$ represent how fire affects tree demography, via forest tree mortality and via the maturation of saplings into savanna trees, and are crucial to the emergence of savanna and forest as alternative stable states in the model. In our theoretical analysis, we assume that $\phi$ and $\omega$ are Heaviside step functions, retaining the key qualitative properties of the fire spread process and simplifying our stability calculations. However, $\phi$ and $\omega$ have smooth, yet sharp, sigmoidal profiles for all numerical investigations (see Table \ref{table.SL_parameters}).}

In recent work, the authors have extended the SL model to a spatially explicit setting by considering interacting particle systems based on the interaction rules outlined above and proving the convergence of these processes to more tractable mean-field limiting processes \cite{patterson2019probabilistic}. The \edit{distribution} of the mean-field limiting processes is governed by the following system of nonlinear integro-differential equations:
\begin{subequations}\label{SL_integral}
	\begin{alignat}{2}
		\partial_t G(x,t) &= \mu S + \nu T + \phi\left(\int_\Omega w(x-y) G(y,t)\,dy \right)F  - \alpha G \int_\Omega J_F(x-y)F(y,t)\,dy \nonumber\\
		&\quad - \beta \,G \int_\Omega J_T(x-y)T(y,t)\,dy, \\
		\partial_t S(x,t) &= -\mu S - \omega\left(\int_\Omega w(x-y)G(y,t)\,dy \right) S - \alpha S \int_\Omega J_F(x-y)F(y,t)\,dy \nonumber\\ &\quad + \beta \,G  \int_\Omega J_T(x-y)T(y,t)\,dy,\\
		\partial_t T(x,t) &= -\nu T + \omega\left( \int_\Omega w(x-y)G(y,t)\,dy \right) S - \alpha T  \int_\Omega J_F(x-y)F(y,t)\,dy, \\
		\partial_t F(x,t) &= \alpha [G + S + T] \int_\Omega J_F(x-y)F(y,t)\,dy - \phi\left(\int_\Omega w(x-y) G(y,t)\,dy\right)F,
	\end{alignat}
\end{subequations}
for each $(x,t) \in \Omega \times \mathbb{R}^+$ for some $\Omega \subset \mathbb{R}^2$ with $\alpha$, $\beta$, $\mu$, and $\nu$ positive constants, $J_F,\, J_T,\,w \in L^1(\Omega;\mathbb{R}_+)$ and $\phi,\, \omega\in C(\mathbb{R}_+;\mathbb{R}_+)$. In our analysis, we will take $\Omega=\mathbb{R}$, but when the system is considered on a compact domain for numerical experiments, we employ periodic boundary conditions. Since the system of equations given by \eqref{SL_integral} describes the evolution of probability densities we retain a normalization condition similar to the space filling constraint for the nonspatial model \eqref{SL_ode}. In particular, 
\begin{equation}\label{eq.normalization}
	G(x,t) + S(x,t) + T(x,t) + F(x,t) = 1 \mbox{ for each } (x,t)\in \Omega\times \mathbb{R}^+.
\end{equation}
The kernel function $w$ measures the ability of fire to spread spatially from a point that is already burning. The constants $\alpha$ and $\beta$ account for the strength of forest-tree and savanna-tree invasion via seed dispersal, with the spatial distribution of these seeds captured by the kernels $J_F$ and $J_T$. The inclusion of nonlocal or long-range interactions is considered most appropriate for spatial vegetation models as dispersal of seeds is often long range or even heavy-tailed~\cite{nathan2012dispersal,thompson2008plant}. The spatial interaction (fire spread and seed dispersal) are assumed isotropic so all kernels are of convolution type and the model \eqref{SL_integral} is thus posed on a homogeneous spatial domain. In all numerical results, we use zero mean Gaussian kernels with different standard deviations (to reflect the relative length scales of the different spatial process). In particular, we have
\[
\mathcal{G}(x,\sigma) := \frac{1}{\sqrt{2\pi \sigma^2}}e^{-x^2/2\sigma^2}, \quad\sigma >0,\quad   x \in \Omega,
\]
with $w(x) = \mathcal{G}(x,\sigma_W)$, $J_T(x) = \mathcal{G}(x,\sigma_T)$, and $J_F(x) = \mathcal{G}(x,\sigma_F)$. Our theoretical results hold for a broad class of kernels obeying some mild assumptions (see below).

\begin{table}
	\centering
	\caption{Summary of parameters and their interpretations for the SL model}
	\begin{tabular}{@{}lcc@{}}
		\toprule
		Ecological interpretation  & Expression & Default numerical value \\
		\midrule
		Forest tree birth rate & $\alpha$ & - \\   
		Savanna saplings birth rate & $\beta$ & -  \\   
		Savanna sapling-to-adult recruitment rate & $\omega(\cdot)$ & $\omega(G) = \omega_0 + \frac{\omega_1-\omega_0}{1 + e^{-(G-\theta_1)/s_1}}$ \\   
		&  & $\omega_0=0.9$, $\omega_1 = 0.4$,  \\ 
		& & $\theta_1 = 0.4$, $s_1 = 0.01$ \\
		Forest tree mortality rate  & $\phi(\cdot)$ & $\phi(G) = \phi_0 + \frac{\phi_1-\phi_0}{1 + e^{-(G-\theta_2)/s_2}} $  \\
		&  & $\phi_0=0.1$, $\phi_1 = 0.9$,   \\
		& & $\theta_2 = 0.4$, $s_2 = 0.05$ \\
		Savanna sapling mortality rate & $\mu$ & 0.1  \\
		Adult savanna tree mortality rate & $\nu$ & 0.05 \\
		\bottomrule
	\end{tabular}
	\label{table.SL_parameters}
\end{table}

\begin{figure}[h]
	\centering
	\includegraphics[width=0.9\textwidth]{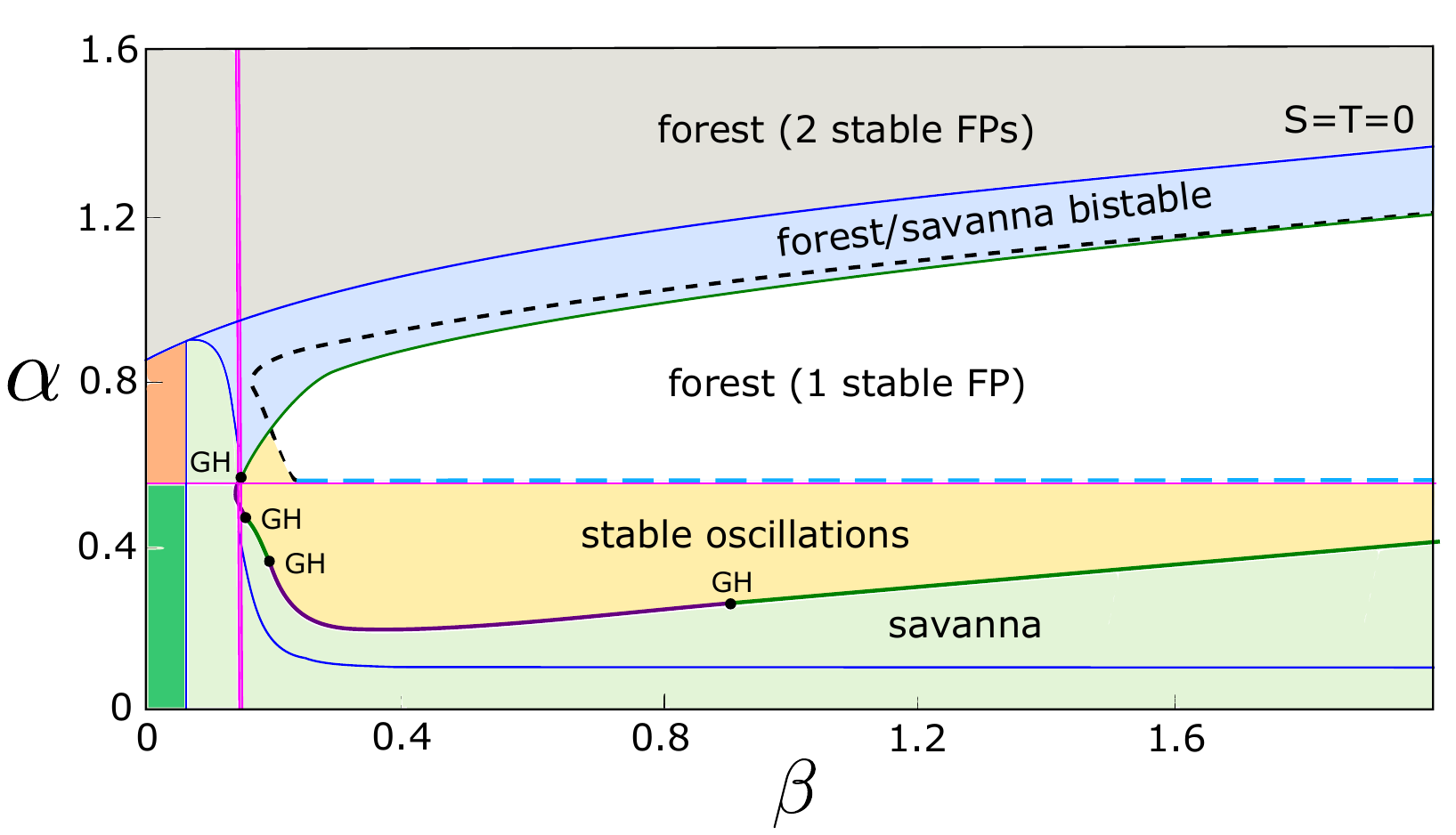}
	\caption{\edit{Two-parameter bifurcation diagram in $\alpha$ (forest tree birth rate) and $\beta$ (savanna tree birth rate) for the nonspatial Staver-Levin model \eqref{SL_ode}. Transcritical bifurcation curves in blue, saddle node curves in magenta, supercritical Hopf curves in purple and subcritical Hopf curves in dark green. Points labeled GH denote codimension 2 Bautin (or Generalized Hopf) bifurcation points at which the Hopf bifurcations change criticality. The all-grass state is the only stable state in the dark green shaded region; the all-grass state and a forest dominated state are bistable in the orange shaded region.}}\label{fig.codim2_nonspatial}
\end{figure}

Figure \ref{fig.codim2_nonspatial} shows a \edit{two-parameter} bifurcation diagram for the nonspatial system \eqref{SL_ode} with the forest tree birth rate, $\alpha$, and the savanna tree birth rate, $\beta$, chosen as bifurcation parameters. Most parameter regimes have one or more stable equilibrium solutions, ranging from all-grass states to savanna and forest states. Moreover, there is significant multistability for large regions of the parameter space and stable oscillations are also possible in the nonspatial model (yellow region in Figure \ref{fig.codim2_nonspatial}). The interested reader may consult \cite{touboul2018complex} for a more detailed bifurcation analysis of the nonspatial SL model as a function of all system parameters, but Figure \ref{fig.codim2_nonspatial} provides a representative summary of the possible dynamics.

\subsection{Pattern formation in the spatially extended SL Model}\label{sec_pattern}
A key goal of the present work is to explore the dynamics of the spatially extended SL model given by \eqref{SL_integral}\edit{,} and a natural first step is to understand the stability of the homogeneous equilibria shown in Figure \ref{fig.codim2_nonspatial} in the presence of spatial interactions.  If $\Omega=\mathbb{R}$ and the kernels $J_F$, $J_T$ and $w$ are appropriately normalized in the spatial model, a steady-state solution to \eqref{SL_ode} is also a solution to \eqref{SL_integral}; it is then evidently of interest to determine if a stable steady-state of \eqref{SL_ode} remains stable in \eqref{SL_integral}. We study the local stability of an arbitrary homogeneous steady-state by linearizing the system about this fixed point and determining whether or not a perturbation about this solution will be damped or amplified by the system dynamics. In particular, we decompose the perturbation over an appropriate Fourier basis; let $\xi \in\mathbb{R}$ denote the wave number (Fourier variable) and suppose $\lambda_{i,\xi}$ for $i=1,...,4$ denotes the eigenvalues associated to the wave number $\xi$ of the linearized system about some homogeneous steady-state. The steady-state solution will be linearly unstable if for some $i=1,...,4$:
\begin{equation}\label{eq.turing_bifurcation}
	\mbox{there exists }\lambda_{i,\xi} \mbox{ such that }\Re{(\lambda_{i,\xi})} > 0 \mbox{ for some }\xi \neq 0,
\end{equation}
where $\Re(\lambda)$ denotes the real part of $\lambda$. Since the homogeneous steady-state was assumed stable in the nonspatial system \eqref{SL_ode}, we have $\lambda_{i,0}<0$ for $i=1,...,4$ and to avoid degeneracy (blow-up of solutions) we must also have that $\limsup_{\lvert\xi\rvert \to\infty}\Re(\lambda_{i,\xi}) <0$. Under these conditions, we expect a branch of stable heterogeneous solutions to emerge as the homogeneous steady-state loses stability, typically referred to as a pattern-forming or symmetry-breaking instability. Since its introduction by Turing \cite{turing1990chemical}, this type of mechanism has found applications in a myriad of fields, including neuroscience, ecology, material science, and various branches of biology, chemistry and physics \cite{cross1993pattern,ermentrout1991stripes,meinhardt1982models}. In particular, there is an extensive literature in ecology on \edit{spatial} patterning in arid ecosystems and associated modeling efforts to explain these patterns mechanistically \cite{gandhi2019vegetation,gowda2014transitions,lefever1997origin,meron2012pattern,von2001diversity}. Here we investigate if the spatial SL model, which is intended as a model for ecosystems at intermediate rainfall levels, also predicts patterned or heterogeneous vegetation distributions in the absence of any externally imposed spatially heterogeneous structure.

We choose $\Omega=\mathbb{R}$ for ease of exposition, but the calculations and conclusions are analogous for other standard domains of interest, for example, $\Omega=\mathbb{R}^2$ and compact subsets of $\mathbb{R}^2$ with periodic boundary conditions. Furthermore, we restrict attention to a class of kernels satisfying the following assumptions:
\begin{enumerate}
	\item[(H1.)] each kernel $J \in L^1(\mathbb{R};\mathbb{R}^+)$ obeys
	$
	\int_{\mathbb{R}}J(x)\,dx = 1,
	$
	\item[(H2.)] $J$ is an even function.
\end{enumerate} 
The normalization condition in (H1.) serves to separate the intensity of spatial processes (such as seed dispersal and fire spread) from their dispersion or variance, and also guarantees that equilibrium solutions of the nonspatial model \eqref{SL_ode} are spatially homogeneous solutions of \eqref{SL_integral}. Condition (H2.) preserves the isotropy of the problem, i.e. we do not consider prevailing wind effects, sloped terrain or other environmental features which might bias the directionality of the fire or seed dispersal processes. For the fire threshold functions, $\phi$ and $\omega$, we assume that each function is a scaled and shifted version of the Heaviside step function, i.e.
\begin{enumerate}
	\item[(H3.)] For some threshold parameter $\theta_1 \in (0,1)$ and level parameters $\omega_0,\,\omega_1$ such that $0 < \omega_1 < \omega_0$,
	\[
	\omega(x) = \begin{cases}
		\omega_0, \quad &x\in[0,\,\theta_1), \\
		\omega_1, &x\in[\theta_1,\,1].
	\end{cases}
	\]
	Similarly, for some $\theta_2 \in (0,1)$ and level parameters $\phi_0,\,\phi_1$ such that $0 < \phi_0 < \phi_1$,
	\[
	\phi(x) = \begin{cases}
		\phi_0, \quad &x\in[0,\,\theta_2), \\
		\phi_1, &x\in[\theta_2,\,1].
	\end{cases}
	\]
\end{enumerate}
In particular, (H3.) implies that $\phi'(x) = \omega'(x) = 0$ for almost every $x\in(0,1)$.

Carrying out the requisite linear stability analysis under the assumptions above yields the following result; the supporting calculations are deferred to Appendix \ref{sec.no_patterns_appendix}.

\begin{proposition}\label{prop.no_patterns}
	If $(\bar{G},\bar{S},\bar{T},\bar{F})$ is a (locally asymptotically) stable equilibrium of the nonspatial SL model given by \eqref{SL_ode}, then it is also a stable spatially homogeneous equilibrium of the spatially extended SL model given by \eqref{SL_integral} with $\Omega=\mathbb{R}$ with kernels obeying (H1.-H2.), and $\omega$ and $\phi$ obeying (H3.).
\end{proposition}

Proposition \ref{prop.no_patterns} states that a stable steady-state solution to the nonspatial model will be locally asymptotically stable in the spatially extended Staver--Levin model given by \eqref{SL_integral}, regardless of the particular distributions chosen for the seed dispersal kernels of forest and savanna trees, and irrespective of the fire spread kernel (as long as they obey (A1.-A2.)). This result encompasses all symmetric spatial kernels that are probability density functions, regardless of the distribution of their mass, or whether they are thin or heavy-tailed. Hence, on a homogeneous spatial domain, stable solutions of the ODE model \eqref{SL_ode} are likely to dominate the dynamics. Moreoever, extensive numerical simulations of the spatial model \eqref{SL_integral} on a homogeneous domain for a range of parameters and initial conditions did not reveal the emergence of any \emph{stable} nonhomogeneous solutions.

Many standard pattern formation paradigms, going back even to the seminal work of Turing~\cite{turing1990chemical}, have the feature of a long-range inhibitory process and a short-range excitation (activation) process (see also \cite{amari1977dynamics}). In the present context, seed dispersal, a long-range excitatory process, and fire spread, a relatively shorter-range inhibitory process, are the spatial processes considered and hence we may have intuited the inability of the system to form patterns, as elucidated in Proposition \ref{prop.no_patterns}. However, Proposition \ref{prop.no_patterns} in fact gives the stronger conclusion that pattern formation can be ruled out even without the need to specify the structure of the kernels, and hence the relative spatial scales of the processes (since only (H1.-H2) are assumed and this does not specify which processes are short or long range). The reason that spatial instabilities are not present, even if one were to unrealistically assume that fire was longer range than seed dispersal,  is that the spatial impact of fire is essentially a higher order effect due to the structure of our model. Although fire does provide an inhibitory interaction that limits tree growth, its (nonlocal) impact is neglected upon linearization and thus is not sufficiently strong to induce an instability. This is due in particular to the steep nature of the sigmoidal fire onset functions $\phi$ and $\omega$, which both have the property that their derivatives are almost everywhere zero (see Appendix \ref{sec.no_patterns_appendix} for further details). We can therefore conclude that our model must take into account additional spatial processes that limit (inhibit) tree growth to make pattern formation possible via a Turing-type mechanism.

\begin{remark}
	\edit{In numerical experiments with fire threshold functions $\phi$ and $\omega$ are smooth steep sigmoids, we did not observe any pattern forming regions apart from those predicted by the linear stability analysis with Heaviside fire threshold functions.}
\end{remark}

\section{Resource limitation effects}\label{sec_resource_limitation}
In many spatially extended ecological models resource limitation is an essential ingredient for the existence of pattern-forming instabilities~\cite{gandhi2019vegetation,gowda2014transitions,lefever1997origin,meron2012pattern,von2001diversity}. Pattern formation in nutrient deficient savanna ecosystems has been studied from both empirical and theoretical perspectives \cite{belsky1994influences,lejeune2002localized} with heterogeneous patterns reported as being particularly common at the interface of savannas and tropical forests~\cite{rietkerk2008regular}. Empirical studies support the hypothesis that forests and savannas are alternative stable states at intermediate mean annual rainfall (MAR) in the tropics~\cite{staver2011global}, and this potential for bistability is well captured in the standard SL model interactions (see \eqref{SL_ode} and Figure \ref{fig.codim2_nonspatial}). However, closed canopy forest ecosystems (i.e. steady states with high values of $F$) should be more favored, even within the bistable range of MAR, as rainfall increases (or the terrain becomes more hospitable with regard to the availability of other resources). This effect is particularly important if we are interested in modeling the forest-savanna transition as it will have strong implications for estimates on the resilience of forest or savanna ecosystems, as well as their abilities to potentially invade one another (cf. \cite{wuyts2019tropical}); introducing resource limitation to the model serves to account for this mediation of the competition between the vegetation types. With regard to the resources necessary to support savanna or forest trees, we have in mind water and soil nutrients, such as nitrogen, which forest trees typically require in more abundance than savanna trees, saplings or grass.  

\begin{figure}[h]
	\centering
	\includegraphics[width=0.99\textwidth]{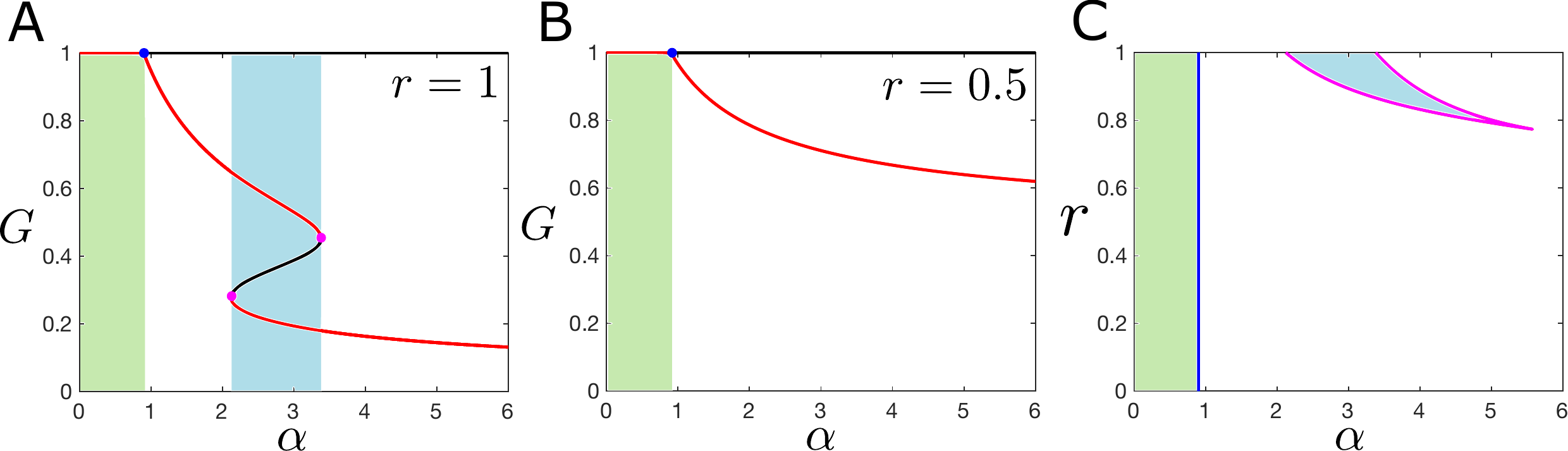}
	\caption{\textbf{A/B:} \edit{One-parameter} bifurcation diagrams for the resource limited model for various values of the resource constraint $r$; stable fixed points in red and unstable fixed points in black. \textbf{C:} \edit{Two-parameter} bifurcation diagram in $r$ and $\alpha$ \edit{(forest tree birth rate)}; curve of transcritical bifurcations in blue, curves of saddle node bifurcations in magenta.}\label{fig.density_bifurcations}
\end{figure}

Consider the grass-forest SL model (ignoring saplings and savannas trees for now) without spatial extent and add a resource limitation effect on forest trees to obtain the following nonspatial model:
\begin{align}\label{GF_MF}
	\dot{G} &= \phi(G)\,F - \alpha\, G\,F \left( 1 - F/r \right), \\
	\dot{F} &= \alpha \,G\,F \left( 1 - F/r  \right) - \phi(G)\,F,
\end{align}
where $r \in [0,1]$ denotes the resource constraint parameter and $G+F=1$. In Figure \ref{fig.density_bifurcations}, we have a numerical bifurcation analysis of system \eqref{GF_MF}; the system exhibits either a stable all-grass solution (green region), bistability between a grassland and a forest dominated solution (blue region), or is monostable (white region) with a solution which contains both grass and forest in proportions that depend on both $\alpha$ and $r$. The regions in which each of these possibilities occurs can be viewed in $\alpha$-$r$ space in the \edit{two-parameter} bifurcation diagram shown in Figure \ref{fig.density_bifurcations} C. 
We now evaluate the stability of each of these steady-state solutions in a spatially extended model. Analogous to the nonlocal spatial extensions shown before, the spatially extended version of \eqref{GF_MF} is given by the following system of integro-differential equations:
\begin{subequations}\label{eq.GF_IDE}
	\begin{alignat}{2}
		\partial_t G(x,t) &= F(x,t)\,\phi\left( \int_\Omega w(x-y)G(y,t)\,dy \right) \\
		&\quad- \alpha \,G(x,t)\int_\Omega J_F(x-y)F(y,t)\,dy \left( 1 - \frac{1}{r}\int_\Omega R(x-y)F(y,t)\,dy   \right), \\
		\partial_t F(x,t) &= \alpha \,G(x,t)\int_\Omega J_F(x-y)F(y,t)\,dy \left( 1 - \frac{1}{r}\int_\Omega R(x-y)F(y,t)\,dy   \right) \\
		&\quad - F(x,t)\,\phi\left( \int_\Omega w(x-y)G(y,t)\,dy \right),\\
		1 &= G(x,t) + F(x,t), \quad \mbox{for each } (x,t)\in\Omega \times \mathbb{R}_+,
	\end{alignat}
\end{subequations}
where the kernel $R$ is a probability density function summarizing the degree of competition that trees exert on one another at various distances. We assume that $R$ obeys (A1.-A2.) and once more fix $\Omega = \mathbb{R}$ for definiteness, but the following calculations and conclusions can of course be extended to more general domains. If $R$ is a Dirac delta distribution, then the resource constraint is analogous to that of the classic Fisher-Kolmogorov equation. Since $F(x,t) = 1 - G(x,t)$ for each $(x,t)\in\mathbb{R} \times \mathbb{R}_+$, the grass-forest system reduces to the single integro-differential equation:
\begin{multline}\label{SL_GF}
	\partial_t G(x,t) = \left(1 - G(x,t)\right)\phi\left( \int_{\mathbb{R}} w(x-y)G(y,t)\,dy \right)\\
	- \alpha G(x,t)\left(1 - \int_{\mathbb{R}} J_F(x-y)G(y,t)\,dy\right) \left( 1 - \frac{1}{r} + \frac{1}{r}\int_{\mathbb{R}} R(x-y)G(y,t)\,dy  \right).
\end{multline}
The grass-forest SL model with resource limitation on forest tree growth given by \eqref{SL_GF} can exhibit pattern-forming instabilities and, via numerical bifurcation analysis, we can further study the structure of these bifurcations and their implications for the ecosystems under consideration. When the resource limitation on forest trees is added to the 4-species model, the only possible spatially induced bifurcations of equilibria are those coming from the forest-grass subsystem (see Section \ref{sec_full_model_patterns} for details); hence the forthcoming detailed analysis of the forest-grass subsystem allows us to understand the source of instabilities in the simplest setting possible. Our analysis shows that threshold fire feedback or disturbance is compatible with pattern formation in resource limited biomes and that there are multiple different pattern formation paradigms in different parameter regimes of $\alpha$-$r$-space and for different relative spatial scales of seed dispersal and resource competition. In particular, we find several different bistable regimes resulting from backward (subcritical) bifurcations of the homogeneous steady states.

\subsection{Stability of steady-state solutions with resource limitation}\label{sec.grass_forest_RL}
We introduce one additional assumption in order to classify the parameter space into regions that can or cannot exhibit pattern formation in the system \eqref{SL_GF}. The key condition for our calculations will be whether or not the Fourier transforms of the kernel functions are nonnegative or not. A simple sufficient condition for a dispersal kernel $J$ to have a nonnegative Fourier transform is that: 
\begin{enumerate}
	\item[(H4.)] $J \in L^1(\mathbb{R};\mathbb{R}^+)$ attains its maximum value at zero.
\end{enumerate}
Hypothesis (H4.) captures virtually all of the kernels typically used in vegetation modeling, including the Gaussian, exponential, and Cauchy kernels. A necessary and sufficient condition for a function to have a nonnegative Fourier transform is that the kernel be positive definite \cite{bochner1959lectures}:
\begin{definition}[Positive definite function]
	A function $f:\mathbb{R}\mapsto\mathbb{C}$ is positive definite if for any real numbers $x_1,\dots,x_n$ the $n\times n$ matrix $A$ whose entries are given by 
	$
	A_{i,j} = f(x_i-x_j)
	$
	is positive semi-definite.
\end{definition}
In most practical cases (H4.) is sufficient and much easier to verify but the forthcoming analysis is unchanged and the results remain valid for the more general class of positive definite kernels. The following result rules out spatially induced interactions destabilizing the all-grass state under the conditions above; the supporting calculations are contained in Appendix \ref{sec_RL_stability_calcs}.

\begin{proposition}\label{prop.no_patterns_all_grass}
	If the all-grass state is a (locally asymptotically) stable equilibrium of the nonspatial resource limited SL model given by \eqref{GF_MF}, then it is also a stable spatially homogeneous equilibrium of the spatially extended resource limited SL model given by \eqref{SL_GF} with $\Omega=\mathbb{R}$ with \emph{positive definite} kernels obeying (H1.), (H2.), and $\omega$ and $\phi$ obeying (H3.).
\end{proposition}

Having ruled out bifurcations of the all-grass solution, we now evaluate the stability \edit{of} mixed equilibria in which both the grass and forest types are present. Define the quantities
\begin{align*}
	A(\alpha,r) &= \phi'(\bar{G})(1-\bar{G}),\quad B(\alpha,r) = -\frac{\alpha}{r} \bar{G}\left(1-\bar{G}\right), \\
	C(\alpha,r) &= \alpha \bar{G}\left(1 - \frac{1-\bar{G}}{r}\right), \quad  D(\alpha,r) = -\alpha \left(1-\bar{G}\right)\left(1 - \frac{1-\bar{G}}{r}\right)- \phi\left(\bar{G}\right).
\end{align*}
The dispersion relation for any homogeneous equilibrium $\bar{G} \in (0,1)$ is then given by
\begin{equation}\label{GF_final_spectrum}
	\lambda_{\xi} = A(\alpha,r)\,\hat{w}(\vert\xi\rvert) + B(\alpha,r) \,\hat{R}\left(\lvert\xi \rvert\right) + C(\alpha,r) \,\hat{J}_F\left( \lvert\xi\rvert \right) + D(\alpha,r),
\end{equation}
where we recall that $\xi$ denotes the wave number (see Appendix \ref{sec_RL_stability_calcs} for the derivation of this expression). For $\bar{G}$ to lose stability and potentially generate new stable heterogeneous solutions the following conditions must hold:
\begin{enumerate}[(i.)]
	\item $\lambda_0 < 0$, i.e. $\bar{G}$ is stable in the ODE \eqref{GF_MF},
	\item there exists $\xi\neq0$ such that $\lambda_\xi > 0$,
	\item $\lambda_\xi < 0$ for all $\lvert\xi\rvert$ sufficiently large. 
\end{enumerate}
We only consider steady-states $\bar{G}$ which are stable for the ODE \eqref{GF_MF} so (i.) is assumed to hold; this is equivalent to asking that
$
A(\alpha,r) + B(\alpha,r) + C(\alpha,r) + D(\alpha,r)<0.
$ 
The Riemann-Lebesgue lemma applies to our kernels by (H1.), so we have
\[
\lim_{\vert\xi\rvert\to\infty} \hat{w}(\lvert\xi\rvert) = \lim_{\lvert\xi\rvert\to\infty} \hat{R}(\lvert\xi\rvert) = \lim_{\lvert\xi\rvert\to\infty} \hat{J}_F(\lvert\xi\rvert) = 0.
\]
For the non-degeneracy condition (iii.) to hold, we must have $ \lim_{\lvert\xi\rvert\to\infty}\lambda_\xi <0$, i.e.
\begin{align}\label{eq.spec_at_infinity}
	\lim_{\lvert\xi\rvert\to\infty}\lambda_\xi &= D(\alpha,r) = -\alpha \left(1 - \frac{1-\bar{G}}{r}\right)- \phi\left(\bar{G}\right) + \alpha\, \bar{G}\left(1 - \frac{1-\bar{G}}{r}\right) < 0 .
\end{align}
Since $\bar{G} \in (0,1)$ obeys 
\begin{equation}\label{eq.SS_RL}
	\phi(\bar{G})\, - \alpha\, \bar{G}\, \left( 1 - \frac{1-\bar{G}}{r} \right) = 0,
\end{equation}
condition \eqref{eq.spec_at_infinity} is equivalent to the inequality
\begin{equation}\label{eq.nondegeneracy}
	1 - \bar{G} = \bar{F} < r,
\end{equation}
which is true for every equilibrium $\bar{G}\in(0,1)$ since the steady state equation \eqref{eq.SS_RL} cannot be satisfied if $1-\bar{G}\geq r$ due to the strict positivity of $\phi$. Therefore the non-degeneracy condition \eqref{eq.spec_at_infinity} is satisfied for all equilibria $\bar{G}\in(0,1)$ and we are guaranteed that condition (iii.) holds. 

Due to \eqref{eq.nondegeneracy}, we have 
\[
A(\alpha, r) \geq 0, \quad B(\alpha,r) < 0, \quad C(\alpha,r)>0, \quad D(\alpha, r) < 0.
\]
It remains to check if condition (ii.) holds. If the kernels are positive definite functions, then
\begin{align}\label{eq.spec_sufficient}
	\sup_{\xi}\lambda_\xi &\leq A(\alpha,r)\,\sup_{\xi}\hat{w}(\lvert\xi\rvert) +  B(\alpha,r) \,\inf_\xi \hat{R}(\lvert\xi\rvert) + C(\alpha,r)\,\sup_\xi\hat{J}_F(\lvert\xi\rvert) + D(\alpha,r)  \nonumber\\
	&= A(\alpha,r) + C(\alpha,r) + D(\alpha,r).
\end{align}
Hence \eqref{eq.spec_sufficient} gives us a necessary condition for (ii.) to hold, namely that 
\[
A(\alpha,r) + C(\alpha,r) + D(\alpha,r) =: N(\alpha,r) > 0,
\]
for a given homogeneous solution $\bar{G}$. Note that since $N(\alpha,r)$ does not depend on the kernel functions, our necessary condition for instabilities does not depend on the nature of the spatial interactions, but only on their overall strengths through the parameters $\alpha$ and $r$. 

\begin{figure}[h] 
	\centering
	\includegraphics[width=0.99\textwidth]{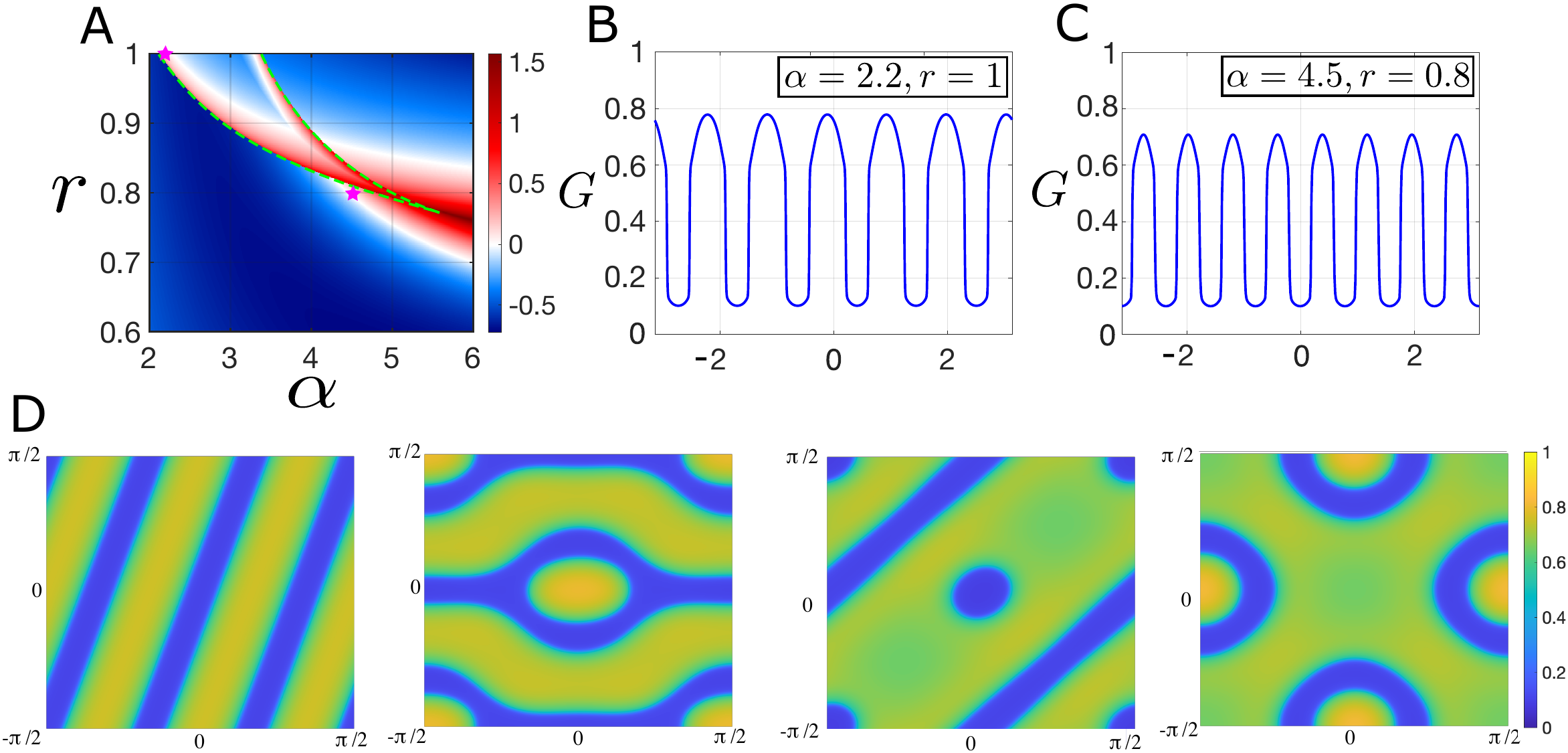}
	\caption{\edit{\textbf{A:} Heatmap of the maximum dispersion relation obtained by linearizing about each (nonspatially) stable homogeneous equilibrium. The saddle-node curves of the nonspatial system are denoted by the green dashed lines and the magenta stars indicate the positions of the parameters for the solutions shown in panels B and C. \textbf{B/C:} Solutions of the spatially extended system with resource limitation on the 1D spatial domain $\Omega = [-\pi,\pi]$ with periodic boundary conditions. \textbf{D:} Selection of solutions on a 2D domain, $\Omega = [-\pi/2,\,\pi/2]\times[-\pi/2,\,\pi/2]$, with periodic boundary conditions ($\alpha = 2.15$ and $r=1$). Other parameters: Gaussian kernels with $\sigma_W=0.01$, $\sigma_F=0.1$ and $\sigma_R=0.4$.}}  \label{fig.density_pattern_reflecting}
\end{figure}

Figure \ref{fig.density_pattern_reflecting} A shows the values of of $\alpha$ and $r$ for which $N(\alpha,r)>0$ for at least one equilibrium that is stable in the ODE model. We only consider values of $\alpha>0.9$ because\edit{,} for $\alpha<0.9$, the all-grass state is the only stable solution of the nonspatial model. Since $N(\alpha,r)$ is a multi-valued function in bistable regions, we take the maximum of $N(\alpha,r)$ across the two stable equilibria in the bistable regions. Instability of the homogeneous solutions is ruled out (regardless of the nature of the spatial interactions) in the white regions which span most of the parameter space. However, there are significant regions highlighted in red where pattern-forming instabilities are possible.  

\begin{figure}[h] 
	\centering
	\includegraphics[width=0.99\textwidth]{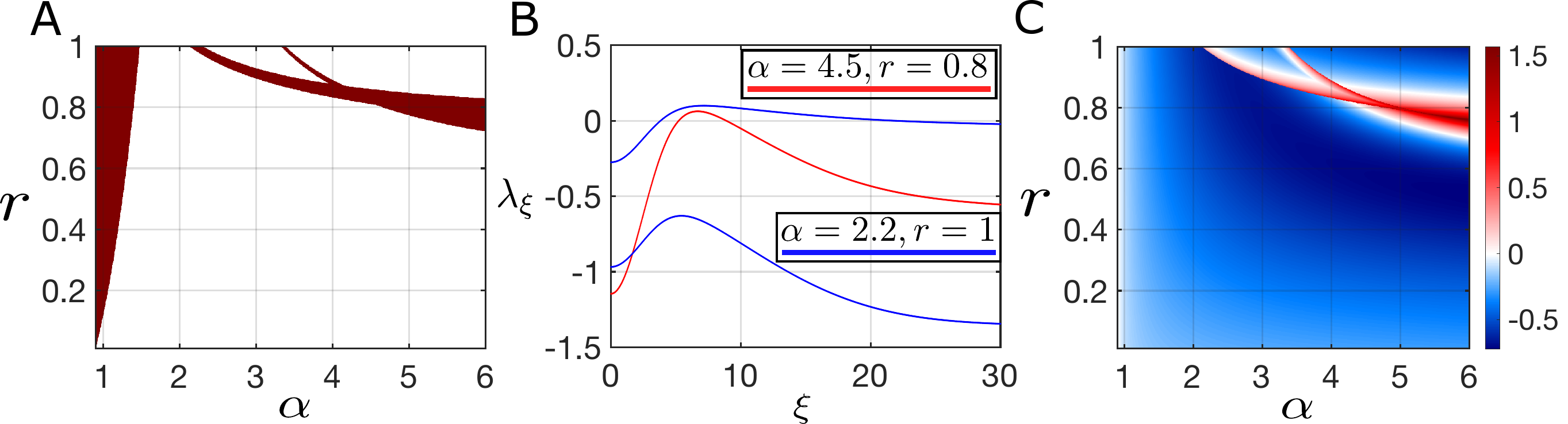}
	\caption{\edit{\textbf{A:} Red shaded regions satisfy the necessary condition for an instability of a spatially homogeneous solution, i.e. $\max\left(N(\alpha,r)\right)>0$, where the max is taken over all stable equilibria of the nonspatial system. \textbf{B:} Plots of the dispersion relations for the spatial model for two parameter sets varying $\alpha$ and $r$ (the system is bistable in one case and monostable in the other). \textbf{C:} Maximum of the principal eigenvalue of the linearized system with red regions denoting pattern-forming parameter regimes. Parameters: Gaussian kernels with $\sigma_W=0.05$, $\sigma_F=0.1$ and $\sigma_R=0.4$.}}  \label{fig.pattern_parameters}
\end{figure}

\edit{In Figure \ref{fig.pattern_parameters} B, we take Gaussian kernels and plot the dispersion relation for the equilibria which are stable for the ODE \eqref{GF_MF} for two sets of parameters, one in the monostable and one in the bistable regime. For the bistable parameter set (blue curves), the dispersion relation of the grass-dominated equilibrium with $\bar{G}\approx 0.62$ is lower blue curve and hence this homogeneous equilibrium does not lose stability in the presence of spatial interactions. However, the dispersion relation of the corresponding forest-dominated equilibrium (with $\bar{G}\approx 0.24$) becomes positive for a range of wave numbers  and hence does lose stability. The dispersion relation in red corresponds to a mixed forest-grass equilibrium in a monostable regime which also loses stability upon the introduction of appropriate spatial interactions.}

Figure \ref{fig.pattern_parameters} C shows the pattern forming instability region across a large region of $\alpha$-$r$ parameter space; this amounts to calculating the dispersion relations at each point and recording the maximum achieved across both equilibria and all wave numbers. The red shaded regions are the so-called Turing space where a stable homogeneous solution has lost stability in the presence of spatial interactions due to a real eigenvalue becoming positive. The plots shown here are with Gaussian kernels for seed dispersal, fire spread and resource limitation. The standard deviation of the Gaussian fire spread kernel ($\sigma_W$) is smaller than that of seed dispersal ($\sigma_F$), which is in turn smaller than that of the resource competition kernel ($\sigma_R$). As we show presently, pattern-forming instabilities persist as the relationships between these parameters are varied over relatively wide ranges and the instabilities do not necessarily depend on having $\sigma_R < \sigma_F$.

\edit{In Figure \ref{fig.density_pattern_reflecting} we confirm the pattern formation predicted in Figure \ref{fig.pattern_parameters} by solving the spatial model with periodic boundary conditions. Figure \ref{fig.density_pattern_reflecting} B and C show stable \emph{heterogeneous} steady state solutions for the system \eqref{eq.GF_IDE} for parameter values predicted to exhibit pattern formation in panel B of Figure \ref{fig.pattern_parameters}. In both cases, a homogeneous solution loses stability and gives birth to a new stable heterogeneous solution; the solution shown in panel B is in the bistable regime and the second homogeneous equilibrium retains stability, while the solution in panel C is not in the bistable regime and appears to be the only stable solution in this parameter regime (up to translation).} Panel D shows some of the complex patterns that arise on a 2 dimensional spatial domain when the homogeneous solution $\bar{G} \approx 0.24$ undergoes a symmetry breaking bifurcation; the rich diversity of possible patterns is due in part to the periodic boundary conditions used in these simulations.

\begin{figure}[h]
	\centering
	\includegraphics[width=0.99\textwidth]{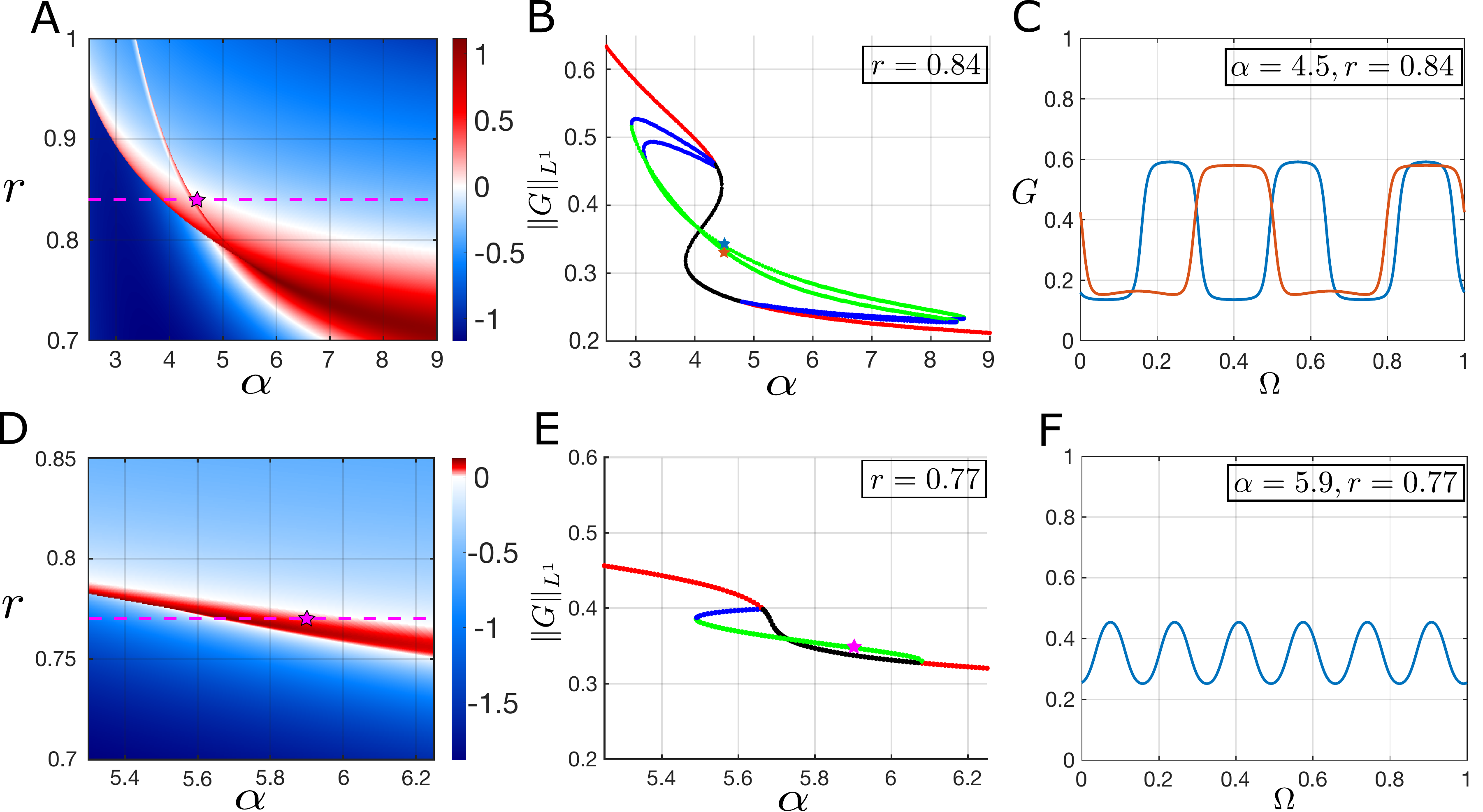}
	\caption{\edit{ \textbf{Top row parameter regime:} $\sigma_F = 0.1$, $\sigma_W = 0.025$, $\sigma_R = 0.15$. \textbf{Bottom row parameter regime:} $\sigma_F = 0.1$, $\sigma_W = 0.025$, $\sigma_R = 0.05$. \textbf{A/D:} Heatmap of the maximum dispersion relation obtained by linearizing about each (nonspatially) stable homogeneous equilibrium. The dashed horizontal magenta line corresponds to the one-parameter bifurcation diagrams in B/E and the magenta star indicates the position of the solutions shown in C/F respectively. \textbf{B/E:} One-parameter bifurcation diagrams for the spatial model varying the forest tree birth rate ($\alpha$); the positions of the solutions shown in panel C are indicated with a star (in panel B the color of the star matches of the color of the solution curve in panel C). \textbf{C/F:} Steady-state profiles of stable heterogeneous solutions with two bistable solutions shown in panel C.}}\label{fig.bifurcations}
\end{figure}

Figure \ref{fig.bifurcations} shows the Turing regions in panels A and D for two different values of $\sigma_R$, the standard deviation of the resource competition kernel, with the kernel much more localized in D than in A, leading to a much smaller Turing region. The Turing regions are considerably larger when $\sigma_R$ is larger (panel A vs panel D), but the situation in panel A is only realistic when we consider resource competition between lower lying vegetation, since this case requires resource competition to be more nonlocal than seed dispersal. \edit{The horizontal dashed magenta line in Figure \ref{fig.bifurcations} A indicates the one-parameter slice of the diagram corresponding to the bifurcation diagram drawn in panel B and the magenta star indicates the position in $\alpha$-$r$ space of the solution shown in panel C.} Panel B of Figure \ref{fig.bifurcations} is a \edit{one-parameter} bifurcation diagram of the spatially extended model \ref{SL_GF} as the seed dispersal rate, $\alpha$, is varied;  the $L^1$ norm of the grass component of the solution shown on the y-axis, stable homogeneous steady states are in red, unstable homogeneous steady states in black, stable heterogeneous steady states and unstable heterogeneous steady states are in green and blue respectively. The homogeneous all-grass solution is not shown but is always unstable for the ranges of $\alpha$ depicted here. 

\edit{In Figure \ref{fig.bifurcations} B, we observe changes in stability in the homogeneous steady states curves around $\alpha\approx 4.35$ and $\alpha\approx4.75$ (at the intersections of the red and black curves). These two bifurcation  points are distinct from the saddle node points of the nonspatial model, indicating that these are instabilities induced by the spatial interactions, matching exactly the predictions of the linear stability analysis in Figure  \ref{fig.bifurcations} A. Numerical continuation reveals that the bifurcations are in fact subcritical and that there are multiple stable heterogeneous solutions. In the bifurcation analysis presented here we show the two stable heterogeneous solution curves which span the widest ranges of $\alpha$ while remaining stable and which also appear to have the largest basins of attraction (based on solving the initial value problem via timestepping). Due to the periodic boundary conditions, there appear to be multiple other unstable solution branches (not pictured), some of which become stable for relatively narrow ranges of the bifurcation parameter. Further analytical work and different methods would be required to understand this behavior systematically, but we emphasize instead the more ecologically relevant aspect of the analysis, i.e. the subcritical nature of the bifurcations and the width of the multistable region. There is a significant region of multistability between the stable heterogeneous solutions (green curves) shown and the stable homogeneous solutions (red curves). Figure \ref{fig.bifurcations} C shows two bistable heterogeneous solution for $\alpha = 4.5$ and $r = 0.84$ with the color of the star in panel B indicating which solution belongs to each stable branch. A two-bump solution curve appears in a saddle-node bifurcation around $\alpha\approx 2.92$, significantly before the linear stability analysis predicts an instability in the homogeneous solutions, and persists until $\alpha \approx 8.44$, when it disappears in another saddle-node bifurcation. The second heterogeneous solution curve is a three-bump solution which similarly appears and disappears in what appear to be saddle-node type bifurcations.}

Figure \ref{fig.bifurcations} D, E and F are analogous to the top row of the same figure in content but focus on a scenario likely to be more realistic for the types of vegetation present in forest-savanna ecosystems, i.e. short scale fire lengthscale ($\sigma_W = 0.025$), intermediate scale resource competition ($\sigma_R = 0.05$) and longer scale seed dispersal ($\sigma_F = 0.1$). As Figure \ref{fig.bifurcations} D shows, the Turing regions are relatively small in this case but the subcritical nature of the bifurcations and the significant early onset of patterning before the bifurcations remains present, as we can see in Figure \ref{fig.bifurcations} E. Indeed, there is still a small region of bistability beyond the point at which the homogeneous branch regains stability around $\alpha\approx6.08$ in panel E, although it is harder to observe here than in panel B. In this case, the heterogeneous branch appears to both gain and lose stability in saddle-node bifurcations, with the unstable branches connecting to the homogeneous branches at the Turing bifurcation points predicted by the linear stability analysis. The solution shown in Figure \ref{fig.bifurcations} F now has a much higher wavelength than before due to the more localized competition (smaller $\sigma_R$).

Our model has three spatial scales: the fire spread scale, the seed dispersal scale and the resource competition scale. We next briefly explore the relationship between the emergence of patterns and the relative values of these scales. Since the dispersion relation that characterizes the onset of pattern formation is not analytically tractable, we resort to numerical analysis to understand the persistence of patterns as we vary the relationships between the spatial scales. As we have throughout, we assume centered Gaussian kernels so that the length scales are summarized by the standard deviation parameters $\sigma_W$, $\sigma_F$ and $\sigma_R$. We take the fire dispersal parameter $\sigma_W = 0.01$ and we fix the seed dispersal parameter $\sigma_F= 0.1$, since forest tree seeds should typically disperse much further than fire embers.

\begin{figure}[h] 
	\centering
	\includegraphics[width=0.99\textwidth]{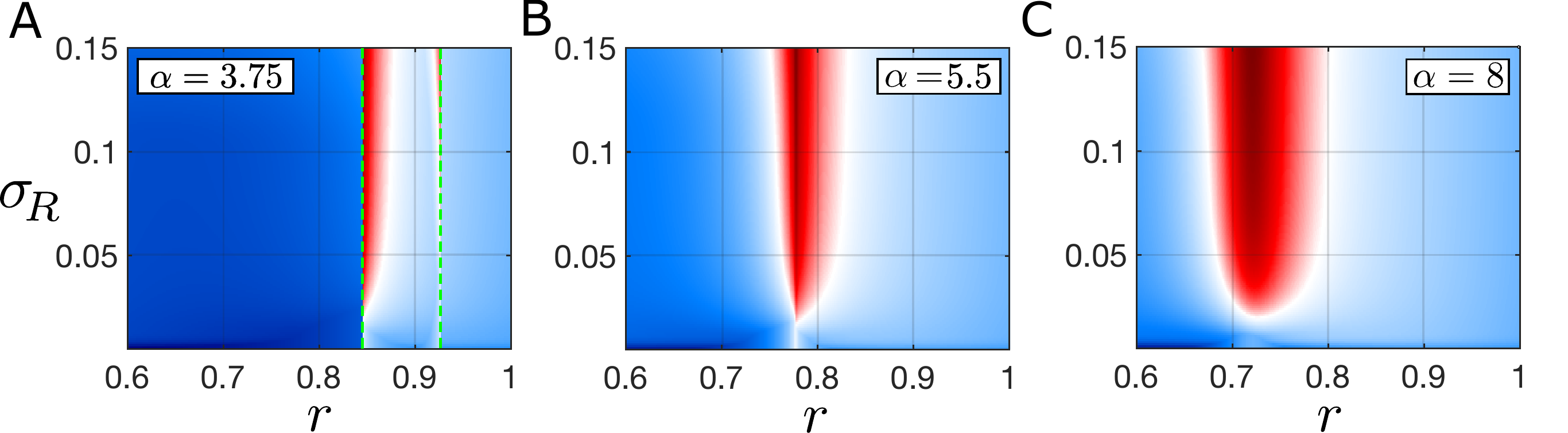}
	\caption{\edit{\textbf{A/B/C:} Heatmaps showing the regions where the dispersion relation is positive for at least one homogeneous equilibrium in the absence of spatial interactions, i.e. the pattern forming region, as a function of $\sigma_R$ (the standard deviation of the resource competition kernel) for different values of $\alpha$ (the forest tree birth rate). The green dashed lines in panel A indicate the position of saddle node bifurcation points (and hence the bistable region) in the non-spatial model. Other parameter values: $\sigma_{F} = 0.1$ and $\sigma_W = 0.01$.}}   \label{fig.length_scale}
\end{figure}

\edit{Figure \ref{fig.length_scale} shows the maximal value of the dispersion relation across all homogeneous steady states of the nonlocal model \eqref{SL_GF} in $r$-$\sigma_R$ space for fixed values of $\sigma_{F} = 0.1$ and $\sigma_W = 0.01$ and various values of $\alpha$. As usual, pattern-forming parameter regions are highlighted in red, with darker red indicating faster growing instabilities. These plots reveal that the pattern-forming instabilities persist for a much larger range of values of $\sigma_R$ than we may have anticipated and are still present even when the $\sigma_R$ is significantly less than $\sigma_F$ ($\sigma_{F} = 0.1$ in all panels). In fact, pattern forming instabilities persist for $\sigma_R$ as low as $\approx 0.02$ across a wide range of values of the resource constraint and forest birth rate parameters ($r$ and $\alpha$ respectively). In panels B and C, the nonspatial system is in monostable parameter regimes, but the system is bistable for part of the range of $r$ values in panel A (between the saddle nodes bifurcations indicated by the vertical dashed green lines).} 

From an ecological perspective, the large range of pattern forming values of $\sigma_R$ is crucial, because it means that patterns may be found even where the distance-scale of resource competition is smaller than that of dispersal; this may occur widely in nature, since resource competition between trees is often considered to be relatively localized compared to dispersal~\cite{nathan2000spatial}. Thus, although the Turing space grows with $\sigma_R$, it is not crucial to our conclusions that $\sigma_R$ be as large as $\sigma_F$, or that resource limitation be of a similar scale to seed dispersal. Other classical models which typically require resource competition to be more nonlocal than growth (or seed dispersal) processes have been employed to explain pattern formation in arid or semi-arid ecosystems with lower lying plant species that disperse very locally~\cite{lefever1997origin}.

\subsection{The four-type SL model with resource limitation}\label{sec_full_model_patterns}
\begin{figure}[h]
	\centering
	\includegraphics[width=0.7\textwidth]{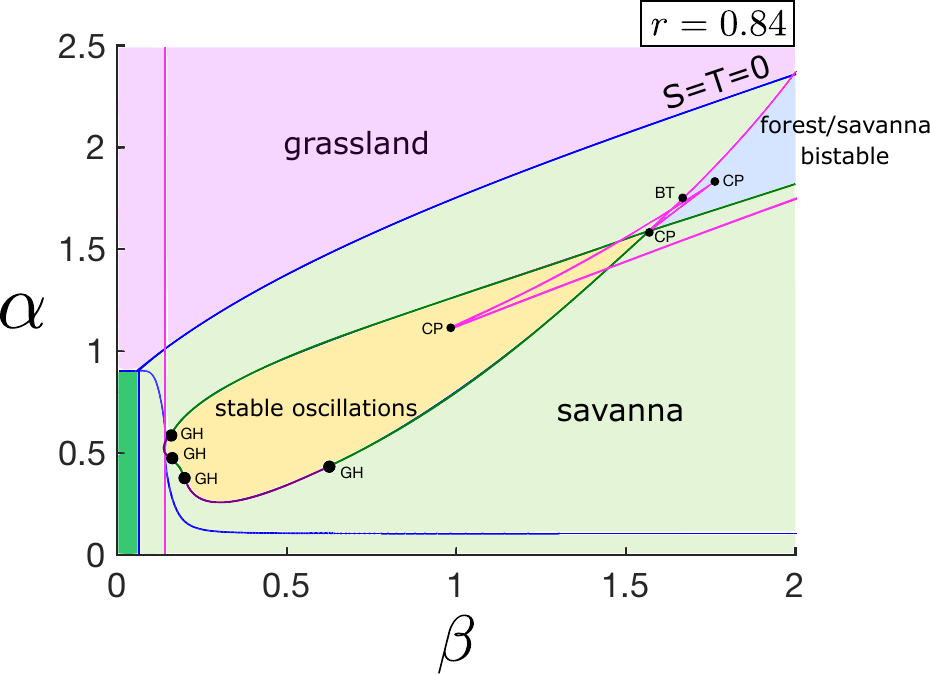}
	\caption{\edit{Two-parameter bifurcation diagram in $\alpha$ (forest tree birth rate) and $\beta$ (savanna tree birth rate) for the nonspatial resource-limited Staver-Levin model \eqref{eq.SL_ODE_RL} with $r = 0.84$. Transcritical bifurcation curves in blue, saddle node curves in magenta, supercritical Hopf curves in purple and subcritical Hopf curves in dark green. Points labeled CP denote codimension 2 Cusp bifurcations and points labeled BT denote codimension 2 Bogdanov-Takens bifurcations. Points labeled GH denote codimension 2 Bautin (or Generalized Hopf) bifurcation points which are points at which the Hopf bifurcations change criticality. The all-grass state is the only stable state in the dark green shaded region (bottom left corner). }}\label{fig.codim2_RL}
\end{figure}

The four-type nonspatial SL model with resource limitation  is given by
\begin{subequations}\label{eq.SL_ODE_RL}
	\begin{alignat}{2}
		\dot{G} &= \mu S + \nu T + \phi(G)F - \alpha G F(1-F/r) - \beta G T, \\
		\dot{S} &= -\mu S - \omega(G) S - \alpha S F(1-F/r) + \beta G T,\\
		\dot{T} &= -\nu T + \omega(G) S - \alpha T F(1-F/r), \\
		\dot{F} &= \alpha (G + S + T)F(1-F/r) - \phi(G)F, \label{eq.RL_noequilibrium}\\
		1 &= G + S + T + F.
	\end{alignat}
\end{subequations}
\edit{Figure \ref{fig.codim2_RL} shows a two-parameter bifurcation diagram for the system \eqref{eq.SL_ODE_RL} as a function of the forest tree birth rate ($\alpha$) and the savanna tree birth rate ($\beta$) for a fixed value of the resource constraint ($r = 0.84$). The presence of the resource constraint pushes the stable forest dominated regions of parameter space to higher values of $\alpha$ when compared to Figure \ref{fig.codim2_nonspatial}, but many of the qualitative features of the model remain similar to the non-resource limited case. We neglect higher values of $\alpha$ in Figure \ref{fig.codim2_RL} since the dynamics become confined to the forest-grass subsystem which was studied extensively in Section \ref{sec.grass_forest_RL}. Stable oscillations are still present in the resource-limited model (yellow shaded region), but in the analysis that follows we continue to focus on homogeneous equilibrium solutions and their stability within the spatially extended version of the model. We did not observe a loss of stability of any homogeneous periodic solutions due to the presence of spatial interactions for the kernels and parameters studied here, but this would be an interesting phenomenon to study in more detail and there are techniques available suited to this problem~\cite{rule2011model}.}
\begin{remark}
	\edit{Figure~\ref{fig.codim2_RL} shows a number of bifurcations around $\alpha \approx 1.65$ and $\beta \approx 1.65$. While bifurcation curves and codimension two points may appear to overlap, no additional degeneracy was observed in the bifurcations. Stable oscillations disappear in this region through a complex sequence of homoclinic and fold of limit cycles bifurcations. While this phenomenon is not of practical interest for this paper, especially since it is confined in a small region of the parameter space, the geometry of the flow in this region is complex and potentially of independent interest.}
\end{remark}
The dynamics of the spatial version of the resource limited model are governed by the following system of equations:
\begin{subequations}\label{eq.SL_integral_RL}
	\begin{alignat}{2}
		\partial_t G(x,t) &= \mu S + \nu T + \phi\left(\int_\Omega w(x-y) G(y,t)\,dy \right)F - \beta \,G \int_\Omega J_T(x-y)T(y,t)\,dy  \nonumber\\
		&- \alpha\, G \left(\int_\Omega J_F(x-y)F(y,t)\,dy\right)\left( 1 - \frac{1}{r}\int_\Omega R(x-y)F(y,t)\,dy \right) , \\
		\partial_t S(x,t) &= -\mu S - \omega\left(\int_\Omega w(x-y)G(y,t)\,dy \right) S + \beta \,G  \int_\Omega J_T(x-y)T(y,t)\,dy  \nonumber\\ &- \alpha S \left(\int_\Omega J_F(x-y)F(y,t)\,dy\right)\left( 1 - \frac{1}{r}\int_\Omega R(x-y)F(y,t)\,dy \right) ,\\
		\partial_t T(x,t) &= -\nu T + \omega\left( \int_\Omega w(x-y)G(y,t)\,dy \right) S \nonumber\\ &\quad- \alpha T  \left(\int_\Omega J_F(x-y)F(y,t)\,dy\right)\left( 1 - \frac{1}{r}\int_\Omega R(x-y)F(y,t)\,dy \right), \\
		\partial_t F(x,t) &= \alpha [G + S + T] \left(\int_\Omega J_F(x-y)F(y,t)\,dy\right)\left( 1 - \frac{1}{r}\int_\Omega R(x-y)F(y,t)\,dy \right) \nonumber\\  &\quad- \phi\left(\int_\Omega w(x-y) G(y,t)\,dy\right)F,
	\end{alignat}
\end{subequations}
for each $(x,t) \in\Omega\times\mathbb{R}^+$. 

To investigate instabilities of homogeneous equilibria of \eqref{eq.SL_integral_RL}, consider a stable equilibrium of the nonspatial model given by \eqref{eq.SL_ODE_RL}, $\left(\bar{G},\bar{S},\bar{T},\bar{F}\right)$, which is also a homogenseous equilibrium of \eqref{eq.SL_integral_RL}. Suppose that hypotheses (H1.-H3.) hold and assume that the spatial kernels are positive definite functions. Upon carrying out the standard linearization procedure (see Appendix \ref{sec.no_patterns_appendix} for details),  we are left to study the eigenvalue problem $\lambda_\xi \vec{u} = A_\xi \vec{u}$ for $\vec{u} = ( \hat{g},\,\hat{s}, \,\hat{\tau},\,\hat{f} )$ where $A_\xi$ is given by
\begin{equation}\label{eq.full_DR_RC}
	{\begin{bmatrix}
			-\beta\bar{T} & \mu & \nu-\beta\bar{G}\hat{J}_T & \phi(\bar{G})-\alpha\bar{G}\hat{J}_F \\ -\,\alpha\bar{F}(1-\bar{F}/r)  & & & +\, \alpha\bar{F} \bar{G}(\hat{ J_F} + \hat{ R})/r \\[10pt]
			\beta\,\bar{T} & -\mu-\omega(\bar{G}) & \beta\bar{G}\hat{J}_T & -\alpha \bar{S}\hat{J}_F  \\ & -\alpha\bar{F}(1-\bar{F}/r)  & & + \alpha\bar{F} \bar{S}(\hat{ J_F} + \hat{ R})/r \\[10pt]
			0 & \omega(\bar{G}) & -\nu & -\alpha\bar{T}\hat{J}_F  \\ & & -\alpha\bar{F}(1-\bar{F}/r) & +\, \alpha\bar{F} \bar{T}(\hat{ J_F} + \hat{ R})/r  \\[10pt]
			& & & \alpha(1-\bar{F})\hat{J}_F-\phi(\bar{G})\\ 0 & 0 & 0 & -\,\alpha\bar{F}(1-\bar{F}/r)  \\ & & & -\, \alpha\bar{F}(1-\bar{F})(\hat{ J}_F+\hat{R})/r
	\end{bmatrix}}
\end{equation}
and $\xi$ denotes the wavenumber of the perturbation. Thus, by expanding the determinant of $A_\xi$ along the fourth row, we can read off that the first set of eigenvalues are given by
\begin{align}\label{eq.forest_RL_spectrum}
	\lambda_{\xi,F} 
	&= \alpha(1-\bar{F})\left(1 -\frac{\bar{F}}{r} \right) \hat{J}_F(\xi)  - \frac{\alpha}{r}\,\bar{F}(1-\bar{F})\hat{R}(\xi)  - \alpha\bar{F} \left(1-\frac{\bar{F}}{r}\right)-\phi(\bar{G}).
\end{align}
The expression \eqref{eq.forest_RL_spectrum} is exactly what we previously obtained when studying the forest-grass subsystem under resource limitation (cf. equation \eqref{GF_final_spectrum}), and only relates to interactions and parameters involving forest and grass alone. Hence our previous analysis of \eqref{eq.forest_RL_spectrum} applies once more. In particular, the homogeneous all-grass solution remains stable when it is stable in the nonspatial model and stable patterns can be found in parameter regimes where the forest-grass subsystem is stable with respect to perturbations ($S=T = 0$ for all $x\in\Omega$).

The other components of the linearized spectrum will be given by the eigenvalues of the reduced coefficient matrix:
\begin{equation}\label{eq.reduced_RC}
	{\begin{bmatrix}
			-\beta\bar{T} & \mu & \nu-\beta\bar{G}\hat{J}_T  \\ -\,\alpha\bar{F}(1-\bar{F}/r)  & & \\[10pt]
			\beta\,\bar{T} & -\mu-\omega(\bar{G}) & \beta\bar{G}\hat{J}_T  \\ & -\alpha\bar{F}(1-\bar{F}/r)  & \\[10pt]
			0 & \omega(\bar{G}) & -\nu  \\ & & -\alpha\bar{F}(1-\bar{F}/r) 
	\end{bmatrix}}.
\end{equation}

\noindent\textbf{Case (i.): $\bar{S}=\bar{T} = 0$}

If we consider stability of homogeneous equilibria within the forest-grass subsystem, then $\bar{S}= \bar{T} = 0$ and the reduced coefficient matrix above becomes
\[
{\begin{bmatrix}
		-\alpha\bar{F}(1-\bar{F}/r)  & \mu & \nu-\beta\bar{G}\hat{J}_T  \\[10pt]
		0 & -\mu-\omega(\bar{G}) & \beta\bar{G}\hat{J}_T  \\ & -\alpha\bar{F}(1-\bar{F}/r)  & \\[10pt]
		0 & \omega(\bar{G}) & -\nu  \\ & & -\alpha\bar{F}(1-\bar{F}/r) 
\end{bmatrix}}.
\]
Thus we can read off that the second set of eigenvalues are given by $
\lambda_{\xi,2} = -\alpha\bar{F}\left(1-\bar{F}/r \right) < 0$ for all $\xi\in\mathbb{R}$  but it is constant in $\xi$ and thus not the source of a potentially instability in the homogeneous solution. In fact, $\lambda_{\xi,2}$ is negative by hypothesis since we assume the equilibrium to be stable for the mean-field model. The final two sets of eigenvalues are those of the coefficient matrix 
\[
{\begin{bmatrix}
		-\mu-\omega(\bar{G}) -\alpha\bar{F}(1-\bar{F}/r) & \beta\bar{G}\hat{J}_T    & \\[10pt]
		\omega(\bar{G}) & -\nu-\alpha\bar{F}(1-\bar{F}/r) 
\end{bmatrix}}
\]
If the trace of the matrix above is negative and the determinant positive for all $\xi\in\mathbb{R}$, then there can be no bifurcation arising from these eigenvalues. The trace is negative by inspection (it does not depend on $\xi$) and the determinant is given by
\begin{align*}
	D(\alpha,\mu,\nu) - \beta\, \bar{G}\,\omega\left(\bar{G}\right)\,\hat{J}_T(\xi) &\geq D(\alpha,\mu,\nu) - \beta\, \bar{G}\,\omega\left(\bar{G}\right)\,\hat{J}_T(0) \\ &= D(\alpha,\mu,\nu) - \beta\, \bar{G}\,\omega\left(\bar{G}\right) > 0,
\end{align*}
where 
$D(\alpha,\mu,\nu) = \left(-\mu-\omega(\bar{G}) -\alpha\bar{F}(1-\bar{F}/r) \right) \left( -\nu-\alpha\bar{F}(1-\bar{F}/r)  \right)$. Therefore the forest-grass subsystem solutions can only lose stability if there exists an $\xi>0$ such that $\lambda_{\xi,F} > 0$ where $\lambda_{\xi,F}$ are the eigenvalues associated with the forest/grass types given by \eqref{eq.forest_RL_spectrum} and there will be no other unstable modes arising from the rest of the system. As we have seen before, this typically leads to stable heterogeneous steady states via (subcritical) Turing bifurcations.

\noindent\textbf{Case (ii.): $\bar{S}>0$ or $\bar{T}>0$}

It remains to check whether there could be an instability of a homogeneous steady state in which savanna and saplings are present arising from the eigenvalues associated with the reduced coefficient matrix \eqref{eq.reduced_RC}. By using the third order Routh-Hurwitz criteria on \eqref{eq.reduced_RC}, it can be shown that no eigenvalues of this matrix can have positive real-part when $\bar{F} \leq r$, even when we are not necessarily in the forest-grass subsystem, i.e. $\bar{S}>0$ or $\bar{T}>0$. However, we cannot have a homogeneous equilibrium with $\bar{F}>r$ since it will not be able to satisfying the steady state equation for forest trees (see equation \eqref{eq.RL_noequilibrium}). Therefore pattern-forming bifurcations only arise in the four-type resource limited model given by \eqref{eq.SL_integral_RL} when an eigenvalue $\lambda_{\xi,F}$ from equation \eqref{eq.forest_RL_spectrum} gains positive real-part. Hence we have reduced the analysis of the highly complex dispersion relation associated with the matrix \eqref{eq.full_DR_RC} to the relatively simpler dispersion relation given by \eqref{eq.forest_RL_spectrum}, which we previously studied in detail in Subsection \ref{sec.grass_forest_RL}. 

\section{Discussion and conclusions}\label{sec_conclusions}

We have outlined a spatially explicit modeling framework for savanna-forest ecosystems at intermediate rainfall and shown that this system admits pattern-forming instabilities leading to stable heterogeneous landscape structures. Moreover, this is only possible when we augment the Staver-Levin model with an additional resource constraint on forest tree growth. The lack of patterns without this additional feature is not necessarily unexpected to those familiar with typical pattern-forming motifs in nonlinear PDEs, but does serve to emphasize the complexity inherent in building a realistic spatial mesic savanna model. As our linear stability analysis shows, the spatial inhibition of the fire process is effectively higher order in our model and thus not sufficiently strong to induce instabilities in the homogeneous solutions, regardless of the assumptions on the relative spatial scales of fire spread and seed dispersal. In effect, our addition of a resource limitation term that lowers the growth rate of forest trees adds a strong inhibitory interaction to the model and thus adds a key ingredient of many standard pattern formation paradigms. 

From an ecological perspective, the resource constraint in our model is currently incorporated in a purely phenomenological way, but there are several candidate spatial processes that may limit forest tree growth in practice. Two possible distinct spatial processes generating this effect are nutrient and water competition, with nutrient competition referring to soil nutrients such as nitrogen. Trees in nutrient poor savannas have been empirically observed to exhibit regular spatial patterns~\cite{lejeune2002localized} and patterns are a key feature of hydrologically coupled vegetation models for semi-arid ecosystems~\cite{meron2012pattern2}.  

\edit{Another key spatial process in savanna ecosystems, which remains relatively less explored from a modeling perspective, is herbivory by large mammals~\cite{staver2021past,staver2020seasonal}. In the case of savanna-forest mosaics, browsing by herbivores like elephants~\cite{cardoso2020role} or nyala~\cite{lagendijk2012short} along the forest edge could reduce tree growth, thereby preventing forest patches from expanding into savannas. For herbivory to have the type of effect that would stabilize a savanna-forest mosaic, browsing would have to intensify with increasing forest patch size; this is counterintuitive since, given the same number of herbivores concentrated on a smaller patch would tend to intensify their effect~\cite{archibald2005shaping,van2003effects}. Consider, however, a savanna-forest mosaic in a system like Lopé National Park in Gabon. Hypothetically, the existence of the savanna-forest mosaic with an extensive network of edges likely allows the persistence of a larger elephant population than would occur otherwise~\cite{cardoso2020role}, which in turn could potentially slow the recruitment of forest trees \cite{terborgh2016megafaunal} and thus help prevent forest from taking over savanna patches. In effect, coupling herbivore population dynamics to landscape vegetation composition could produce emergent dynamics that are currently neglected in most theoretical models for herbivore impacts in savannas. This potential novel mechanism for stabilizing vegetation patterns would benefit from further theoretical (as well as empirical) investigation.}

Numerical simulations and continuation analysis revealed the subcritical nature of the pattern-forming bifurcations in our model, giving rise to the onset of stable heterogeneous solutions well before the spatially homogeneous steady states lose stability. These numerical results emphasize the necessity of going beyond near equilibrium analysis to understand the possible transitions that might occur as systems are subjected to exogenous perturbations (see Figure \ref{fig.bifurcations}). \edit{We showed two heterogeneous solution curves in our continuation results, but there are many more unstable solutions curves not shown, some of which appear to snake back and forth as the bifurcation parameter is varied and may gain stability over relatively narrow parameter ranges. We did not show these curves for visual clarity, but it would be interesting to investigate the nature of this behaviour and to understand to what extent it persists when the symmetry induced by the periodic boundary conditions is broken.}

The main conclusions of the bifurcation analysis are robust to variation in the relative spatial scales in the problem, particularly the length scales of the seed dispersal and resource limitation processes. Although pattern formation occurs over wider ranges of parameter space when resource competition is more nonlocal than seed dispersal, there are still significant regions of parameter space that omit patterns when seed dispersal is considered to be more long range than competition for resources. This is an important consideration for real-world applications as, in contrast with low lying vegetation patterns such as the famous tiger bush~\cite{lefever1997origin}, we expect wind dispersal of forest tree seeds to be the longest range spatial process in our system. From an ecological standpoint, the robust persistence of patterned vegetation in mesic savanna/forest systems (referred to in the empirical literature as a ``savanna-forest mosaic''~\cite{nathan2000spatial}) is also significant, and suggests that the combination of fire, dispersal limitation, and resource constraints may contribute to stabilizing mosaics in intermediate rainfall landscapes.

A central goal for the ecological and mathematical modeling community interested in savanna-forest ecosystems is a unified mathematical model able to explain observed savanna-forest configurations over a broad range of mean annual rainfall (and subject to local topography). Although such a unified forest-savanna model is crucial for prediction and conservation efforts, no such model currently exists in the literature~\cite{rietkerk2021evasion}. Such a unified model would be required to reproduce observed spatial patterns, as shown here and in other papers focusing on drier savannas~\cite{baudena2013complexity,groen2017spatially,tzuk2020role}, and provide realistic ecosystems descriptions by reflecting the diversity of functional types within different tropical ecosystems. We believe the model proposed here is somewhat unique in satisfying both of these criteria while retaining sufficient tractability to allow detailed analysis of solutions. Finally, while this study is limited to a spatially homogeneous domain, and hence most appropriate for modeling at relatively small spatial scales, we note that there is growing acknowledgment in the modeling community of the key role that heterogeneity (or underling spatial pattern of the domain) plays in determining the final spatial patterns that we observe empirically~\cite{bastiaansen2022fragmented,rietkerk2021evasion}. This consideration is particularly relevant in the present context as savannas are subject to significant continental scale rainfall gradients in Africa and Amazonia~\cite{bucini2007continental,wuyts2017amazonian}, and thus we view the present work as a prerequisite step for a more realistic mesic savanna model that would produce patterns via an interplay between emergent effects and externally imposed spatial pattern owing to domain heterogeneity. Significant challenges remain as we progress towards a unified savanna-forest modeling framework that could be empirically validated, and ultimately this is likely to require integrating hydrology and heterogeneity, along with the functional vegetation types and spatial interactions presented here. 

\section*{Acknowledgments}

Denis Patterson and Simon Levin are grateful for the support of NSF DMS-1951358. Carla Staver acknowledges support from NSF DMS-1615531 and Jonathan Touboul acknowledges support from NSF DMS-1951369.

\section*{Data availability}

All codes used to generate the figures and results in this paper are available at \url{https://github.com/patterd2/mesic_savanna_patterns}.

\appendix

\section{Supplementary Information}\label{sec.supplementary}
	\subsection{Stability of homogeneous equilibria in the standard SL model}\label{sec.no_patterns_appendix}
	The following proposition recalls some elementary implications of assumptions (H1.) and (H2.) crucial to the stability calculations which follow.
	\begin{proposition}\label{prop.assumptions}
		If $J \in L^1(\mathbb{R};\mathbb{R}^+)$ obeys (H1.) and (H2.), then
		\begin{enumerate}[(a.)]
			\item $\hat{ J}(0) = 1$ and $\hat{ J}(\xi) \leq 1$ for each $\xi\in\mathbb{R}$,
			\item $\lim_{\vert\xi\rvert\to\infty}\hat{ J}(\xi) = 0$ (Riemann-Lebesgue Lemma),
			\item $\hat{ J}(-\xi) = \hat{ J}(\xi)  =\hat{J}(\lvert\xi\rvert)$ for each $\xi\in\mathbb{R}$,
		\end{enumerate}
		where $\hat{ J}(\xi) = \int_{\mathbb{R}} J(x) e^{i\xi x}\,dx$ denotes the Fourier transform of $J$ for each $\xi\in\mathbb{R}$.
	\end{proposition}
	Suppose $(\bar{G},\bar{S},\bar{T},\bar{F})$ is a stable equilibrium solution of \eqref{SL_ode}. Due to (H1.) $(\bar{G},\bar{S},\bar{T},\bar{F})$ will also be a solution to the system \eqref{SL_GF}. Now consider a perturbation to this equilibrium solution of the form $(\bar{G}+\epsilon g(x,t),\bar{S}+ \epsilon s(x,t),\bar{T}+ \epsilon\tau(x,t),\bar{F}+\epsilon f(x,t))$. Expand the heterogeneous perturbation terms in the Fourier space as follows:
	\begin{align*}
		\bar{G} + \epsilon \,g(x,t) &= \bar{G} + \epsilon\int_{\mathbb{R}} \hat{g}(\xi,t)e^{i\xi x}\,d\xi,\quad \bar{S} + \epsilon \,s(x,t) = \bar{S} + \epsilon\int_{\mathbb{R}} \hat{s}(\xi,t)e^{i\xi x}\,d\xi, \\
		\bar{T} + \epsilon \,\tau(x,t) &= \bar{T} + \epsilon\int_{\mathbb{R}} \hat{\tau}(\xi,t)e^{i\xi x}\,d\xi,\quad \bar{F} + \epsilon \,f(x,t) = \bar{F} + \epsilon\int_{\mathbb{R}} \hat{f}(\xi,t)e^{i\xi x}\,d\xi,
	\end{align*}
	where $\hat{g}$ denotes the Fourier transform of the perturbation term $g$ and $
	\hat{g}(\xi,t) = e^{\lambda_\xi t} \hat{g}(\xi,0) \equiv e^{\lambda_\xi t} \hat{g}(\xi)
	$ denotes the amplitude associated to the so-called wave number $\xi\in\mathbb{R}$. Linearizing \eqref{SL_integral} around the equilibrium solution $(\bar{G},\bar{S},\bar{T},\bar{F})$ and truncate at first order in the perturbations to obtain the linearisation of the system about the chosen spatially homogeneous steady-state. After eliminating the exponential prefactors, we see that the eigenvalues $\lambda_\xi$ associated to the wave number $\xi$ obey:
	\begin{align*}
		\lambda_\xi \,\hat{g}(\xi) &= \mu\, \hat{s}(\xi) + \nu\,\hat{ \tau}(\xi) - \alpha\, \bar{G} \,\hat{J}_F(\xi) \,\hat{f}(\xi) + \phi(\bar{G})\,\hat{f}(\xi) -\beta\,\bar{G} \,\hat{J}_T(\xi) \,\hat{\tau }(\xi) \\  &\quad + \phi'(\bar{G})\,\bar{F} \,\hat{w}(\xi)\,\hat{g}(\xi)  -\alpha\, \bar{F}\, \hat{g}(\xi)-\beta\,\bar{T}\,\hat{g}(\xi),\\
		\lambda_\xi \,\hat{ s}(\xi) &= -\mu\, \hat{ s}(\xi) - \omega(\bar{G})\, \hat{ s}(\xi) + \beta\,\bar{T}\,\hat{ g}(\xi) -\omega'(\bar{G})\,\bar{S}\, \hat{ w}(\xi)\,\hat{ g}(\xi) \\ &\quad -\alpha\,\bar{S}\,\hat{ J}_F(\xi)\,\hat{ f}(\xi) + \beta\, \bar{G}\,\hat{J}_T(\xi)\,\hat{\tau}(\xi) - \alpha\, \bar{F} \, \hat{s}(\xi), \\ 
		\lambda_\xi\, \hat{ \tau}(\xi) &= -\alpha\, \bar{F}\,\hat{\tau}(\xi) -\nu\,\hat{\tau}(\xi) +\omega(\bar{G}) \, \hat{ s}(\xi)+\omega'(\bar{G}) \, \bar{S}\, \hat{ w}(\xi) \,\hat{ g}(\xi) -\alpha\, \bar{T}\, \, \hat{ J}_F(\xi) \,\hat{ f}(\xi), \\
		\lambda_\xi\, \hat{f}(\xi) &= \alpha(1-\bar{F} )\,\hat{J}_F(\xi)\,\hat{f}(\xi) -\phi(\bar{G})\,\hat{ f}(\xi)-\alpha \bar{F}\hat{f}(\xi)  -\phi'(\bar{G})\,\bar{F} \,\hat{w}(\xi)\,\hat{g}(\xi).
	\end{align*}
	Thus for each $\xi\in\mathbb{R}$, we have the eigenvalue problem
	\[
	\lambda_\xi\begin{bmatrix}
		\hat{g}\\ \hat{s}\\ \hat{\tau}\\ \hat{f} 
	\end{bmatrix} =
	A
	\begin{bmatrix}
		\hat{g}\\ \hat{s}\\ \hat{\tau}\\ \hat{f} 
	\end{bmatrix},
	\]
	where the coefficient matrix $A$ is given by 
	\[
	\begin{bmatrix}
		\phi'(\bar{G})\bar{F}\hat{w}-\alpha\bar{F}-\beta\bar{T} & \mu & \nu-\beta\bar{G}\hat{J}_T & \phi(\bar{G})-\alpha\bar{G}\hat{J}_F \\
		\beta\bar{T}-\omega'(\bar{G})\bar{S}\hat{w} & -\mu-\alpha\bar{F} -\omega(\bar{G}) & \beta\bar{G}\hat{J}_T & -\alpha \bar{S}\hat{J}_F \\
		\omega'(\bar{G}) \bar{S}\hat{w} & \omega(\bar{G}) & -\nu-\alpha\bar{F} & -\alpha\bar{T}\hat{J}_F \\
		-\phi'(\bar{G})\bar{F}\hat{w} & 0 & 0 & \alpha(1-\bar{F})\hat{J}_F-\alpha\bar{F}-\phi(\bar{G})
	\end{bmatrix}
	\]
	Owing to hypothesis (H3.), $\phi'(\bar{G}) = \omega(\bar{G}) = 0$ for a.e. $\bar{G}$ and hence reduces the coefficient matrix $A$ reduces to
	\[
	\begin{bmatrix}
		-\alpha\bar{F}-\beta\bar{T} & \mu & \nu-\beta\bar{G}\hat{J}_T & \phi(\bar{G})-\alpha\bar{G}\hat{J}_F \\
		\beta\bar{T} & -\mu-\alpha\bar{F} -\omega(\bar{G}) & \beta\bar{G}\hat{J}_T & -\alpha \bar{S}\hat{J}_F \\
		0 & \omega(\bar{G}) & -\nu-\alpha\bar{F} & -\alpha\bar{T}\hat{J}_F \\
		0 & 0 & 0 & \alpha(1-\bar{F})\hat{J}_F-\alpha\bar{F}-\phi(\bar{G})
	\end{bmatrix}
	\]
	At this point, no nonlocal effects of fire remain in the model. Expanding along the fourth row of the matrix above, we immediately identify that the component of the spectrum relating to forest trees can be considered in isolation and is given by
	\begin{equation}\label{F_spectrum}
		\lambda_F\left(\xi\right) = \lambda_F\left(\lvert\xi\rvert\right) = \alpha(1-\bar{F})\hat{J}_F\left(\lvert\xi\rvert\right)-\alpha\bar{F}-\phi(\bar{G}).
	\end{equation}
	This component of the spectrum is stable in the mean--field model if and only if
	\begin{equation}\label{F_stable_MF}
		\alpha(1-\bar{F})-\alpha\bar{F}-\phi(\bar{G}) < 0,
	\end{equation}
	Condition \eqref{F_stable_MF} corresponds to asking that this part of the spectrum is stable for the zero Fourier mode, i.e. $\lambda_F(0)<0$. In order for a Turing bifurcation to arise here we assume that \eqref{F_stable_MF} holds and try to find a (nonzero) value of $\xi$ such that $\lambda_F(\lvert\xi\rvert)>0$. However, by Proposition \ref{prop.assumptions} (a.),
	\begin{align*}
		\sup_{\xi}\lambda_F(\lvert\xi\rvert) &=  \alpha(1-\bar{F})\sup_{\xi}\hat{J}_F\left(\lvert\xi\rvert\right)-\alpha\bar{F}-\phi(\bar{G}) = \alpha(1-\bar{F})\hat{J}_F\left(0 \right)-\alpha\bar{F}-\phi(\bar{G}) \\ &=  \alpha(1-\bar{F})-\alpha\bar{F}-\phi(\bar{G})  + \phi(\bar{G}) < 0.
	\end{align*}
	Hence this component of the spectrum is stable in the spatially extended model whenever it is stable for the mean--field model and therefore cannot destabilize the spatially homogeneous steady-state. 
	
	Next consider the eigenvalues relating to the reduced coefficient matrix, i.e. 
	\[
	\begin{bmatrix}
		-\alpha\bar{F}-\beta\bar{T} & \mu & \nu-\beta\,\bar{G}\,\hat{J}_T  \\
		\beta\bar{T} & -\mu-\alpha\bar{F} -\omega(\bar{G}) & \beta\,\bar{G}\,\hat{J}_T  \\
		0 & \omega(\bar{G}) & -\nu-\alpha\bar{F} 
	\end{bmatrix}
	\]
	The characteristic polynomial for the reduced coefficient matrix above has the form 
	\[
	\lambda_\xi^3 + a_2(\xi) \lambda_\xi^2 + a_1(\xi) \lambda_\xi + a_0(\xi) = 0,
	\]
	for each fixed $\xi\in\mathbb{R}$. For a given $\xi\in\mathbb{R}$, the Routh-Hurwitz criteria state that all roots of the characteristic polynomial have strictly negative real parts if:
	\[
	a_2(\xi)>0,\quad a_0(\xi)>0\quad \mbox{and}\quad a_2(\xi)\,a_1(\xi)>a_0(\xi). 
	\]
	Hence, we must have
	\[
	a_2(0)>0,\quad a_0(0)>0\quad \mbox{and}\quad a_2(0)\,a_1(0)>a_0(0), 
	\]
	since the steady-state solution we are considering is stable for the mean-field system given by \eqref{SL_ode}. In our case, $
	a_2(\xi) = 3\alpha\,\bar{F} +\beta\,\bar{T} + \mu + \nu + \omega\left(\bar{G}\right) = a_2(0) > 0,
	$ so $a_2(\xi)>0$ for all $\xi\in\mathbb{R}$. Thus the first Routh-Hurwitz condition cannot be the source of a potential Turing bifurcation. The coefficient $a_0(\xi)$ is given by 
	\[
	\alpha  \bar{F} (\omega\left(\bar{G}\right) (\alpha  \bar{F}-\beta  \bar{G} \hat{J}_T(\vert\xi\rvert)+\nu +\beta  \bar{T})+(\alpha  \bar{F}+\nu ) (\alpha \bar{F}+\mu +\beta \bar{T})).
	\] 
	The condition $a_0(\xi)>0$ holds if and only if 
	\begin{equation}\label{eq.a0}
		A(\alpha,\beta,\nu,\mu) > \beta  \bar{G}  \bar{F} \omega\left(\bar{G}\right) \,\hat{J}_T(\lvert\xi\rvert) ,
	\end{equation}
	where $
	A(\alpha,\beta,\nu,\mu)  = \bar{F} (\omega\left(\bar{G}\right) (\alpha  \bar{F}+\nu +\beta  \bar{T})+(\alpha  \bar{F}+\nu ) (\alpha \bar{F}+\mu +\beta \bar{T})).
	$ Crucially, $A$ does not depend on the Fourier variable $\xi$. Since $a_0(0)>0$, 
	\[
	A(\alpha,\beta,\nu,\mu)  > \beta  \bar{G}  \bar{F} \omega\left(\bar{G}\right) \geq  \beta  \bar{G}  \bar{F} \omega\left(\bar{G}\right) \,\hat{J}_T(\lvert\xi\rvert)\quad \mbox{ for all }\xi\in\mathbb{R}
	\]
	because the maximum value of $\hat{ J}_T$ occurs at $\xi=0$ by Proposition \ref{prop.assumptions} (a.). Hence the second Routh-Hurwitz condition holds for all $\xi\in\mathbb{R}$. Finally, the expression for $a_1(\xi)$ is given by
	\[
	3 \alpha^2 \bar{F}^2+2 \alpha  \bar{F} (\mu +\nu +\beta  \bar{T}+\omega\left(\bar{G}\right))-\beta  G \hat{J}_T(\vert\xi\rvert) \omega\left(\bar{G}\right)+\nu  (\mu + \omega\left(\bar{G}\right))+\beta  T (\nu + \omega\left(\bar{G}\right)),
	\]
	and hence $a_2(\xi)\,a_1(\xi)>a_0(\xi)$ if and only if
	\begin{equation}\label{eq.RH_ineq}
		B(\alpha,\beta,\nu,\mu) \, ( \omega\left(\bar{G}\right) (2 \alpha  \bar{F}-\beta  \bar{G} \hat{J}_T(\vert\xi\rvert) +\nu +\beta \bar{T}) +(2 \alpha \bar{F}+\nu ) (2 \alpha \bar{F}+\mu +\beta \bar{T}))>0,
	\end{equation}
	where $B(\alpha,\beta,\nu,\mu) = 2 \alpha  \bar{F}+\mu +\nu +\beta  \bar{T}+ \omega\left(\bar{G}\right)$. Rearrangement shows inequality \eqref{eq.RH_ineq} above can be written as 
	\begin{equation*}
		B(\alpha,\beta,\nu,\mu)\,C(\alpha,\beta,\nu,\mu) > D(\alpha,\beta,\nu,\mu)\, \hat{J}_T(\vert\xi\rvert),
	\end{equation*}
	where $
	C(\alpha,\beta,\nu,\mu) =  \omega\left(\bar{G}\right) (2 \alpha  \bar{F} +\nu +\beta \bar{T})+(2 \alpha \bar{F}+\nu ) (2 \alpha \bar{F}+\mu +\beta \bar{T})
	$ and $
	D(\alpha,\beta,\nu,\mu) = \beta \omega\left(\bar{G}\right) \bar{G}  (2 \alpha  \bar{F}+\mu +\nu +\beta  \bar{T}+ \omega\left(\bar{G}\right)).
	$ Since $a_2(0)\,a_1(0)>a_0(0)$ and the maximum of $\hat{ J}_T$ occurs at $\xi=0$,
	\[
	B(\alpha,\beta,\nu,\mu)\,C(\alpha,\beta,\nu,\mu) > D(\alpha,\beta,\nu,\mu) \geq  D(\alpha,\beta,\nu,\mu)\, \hat{J}_T(\lvert\xi\rvert),
	\]
	and the final Routh-Hurwitz criterion thus holds for all $\xi\in\mathbb{R}$. Therefore all eigenvalues $\lambda_\xi$ have negative real part for all $\xi\in\mathbb{R}$ and the steady-state solution is linearly stable in the spatially extended system \eqref{SL_integral}. 
	
	\subsection{Stability of homogeneous equilibria with resource limitation}\label{sec_RL_stability_calcs}
	
	Once more we linearize about a fixed point of the mean--field version of the model, i.e. we consider a stable equilibrium solution $\bar{G}$ of \eqref{GF_MF} and then take a perturbation of this equilibrium solution of the form 
	\begin{equation}\label{eq.pertrub_ansatz}
		G(x,t) = \bar{G} + \epsilon \,g(x,t) = \bar{G} + \epsilon\int_{\mathbb{R}} \hat{g}(\xi,t)e^{i\xi x}\,d\xi,
	\end{equation}
	where $\hat{g}$ denotes the Fourier transform of the perturbation term $g$ and $
	\hat{g}(\xi,t) = e^{\lambda_\xi t} \hat{g}(\xi,0) \equiv e^{\lambda_\xi t} \hat{g}(\xi)
	$ denotes the amplitude associated with the wave number $\xi\in\mathbb{R}$. Inserting the perturbed solution from \eqref{eq.pertrub_ansatz} into \eqref{SL_GF} and discarding second order (and higher) terms in $\epsilon$ shows that to leading order $\lambda_\xi$ obeys:
	\begin{align*}
		\lambda_\xi \,\hat{g}(\xi) &= \phi '\left(\bar{G}\right)\left(1 - \bar{G}\right)\hat{w}(\xi)\,\hat{g}(\xi) - \phi\left(\bar{G}\right)\hat{g}(\xi) - \frac{\alpha}{r}\bar{G}\left(1-\bar{G}\right)\hat{R}(\xi)\hat{g}(\xi) \\ &\quad+ \alpha\, \bar{G}\left( 1 - \frac{1-\bar{G}}{r} \right)\hat{J}_F(\xi)\, \hat{g}(\xi) - \alpha\, \left(1-\bar{G}\right)\left( 1 - \frac{1-\bar{G}}{r}\right)\hat{g}(\xi),
	\end{align*}
	for each $\xi\in\mathbb{R}$. Our kernels are all even functions by assumption, so we obtain the simplified dispersion relation
	\begin{align} \label{GF_spectrum}
		\lambda_\xi &= \phi '\left(\bar{G}\right)\left(1 - \bar{G}\right)\hat{w}\left(\lvert\xi\rvert\right) - \phi\left(\bar{G}\right) - \frac{\alpha}{r}\bar{G}\left(1-\bar{G}\right)\hat{R}\left(\lvert\xi\rvert\right) \nonumber\\ &+ \alpha \bar{G}\left( 1 - \frac{1-\bar{G}}{r} \right)\hat{J}_F \left(\lvert\xi\rvert\right) - \alpha \left(1-\bar{G}\right)\left( 1 - \frac{1-\bar{G}}{r}\right),
	\end{align}
	for each $\xi \in \mathbb{R}$. If we consider the all grass state $\bar{G} = 1$ (which is stable for all values of $\alpha$ before the transcritical bifurcation denoted by the vertical blue line in Figure \ref{fig.density_bifurcations} C), then equation \eqref{GF_spectrum} reduces to $
	\lambda_\xi = -\phi\left(1\right) + \alpha \hat{J}_F\left(\lvert\xi\rvert\right)$ for each $\xi\in\mathbb{R}. 
	$ Since the Fourier transform of a nonnegative function achieves its maximum at zero, 
	$
	\sup_{\xi}\lambda_\xi = -\phi\left(1\right) + \alpha \hat{J}_F\left(0\right) = - \phi(1) + \alpha
	$ and hence the stability criterion for the all grass state is the same as it was for the mean--field model \eqref{SL_ode}. Thus if $\bar{G} = 1$ was stable for the nonspatial model \eqref{GF_MF}, it will remain stable for the spatially extended model given by \eqref{eq.GF_IDE}. This means that in $r$-$\alpha$ space, the all grass steady-state is stable in the spatial model in the green shaded region of parameter space from Figure \ref{fig.density_bifurcations} D if we start sufficiently close to $\bar{G}=1$, regardless of the nature of the spatial interactions (i.e. for any kernels obeying assumptions (H1.) and (H2.)). 

\bibliographystyle{abbrv}
\bibliography{savanna_refs}
\end{document}